\documentclass{IEEEtran}
\usepackage{cite}
\usepackage{amssymb,amsfonts}
\usepackage[intlimits]{amsmath}
\usepackage{algorithmic}
\usepackage{graphicx}
\usepackage{textcomp}
\usepackage{xcolor}
\usepackage{pbalance}
\def\BibTeX{{\rm B\kern-.05em{\sc i\kern-.025em b}\kern-.08em
    T\kern-.1667em\lower.7ex\hbox{E}\kern-.125emX}}
\begin{document}
\title{Geometrical Estimation of Electromagnetic Degrees of Freedom in Reflecting Environments}
\author{Yanwen Chen, \IEEEmembership{Graduate Student Member, IEEE}, Stefano Maci, \IEEEmembership{Fellow, IEEE} and Mats Gustafsson, \IEEEmembership{Senior Member, IEEE}
\thanks{This work was supported in part by the Swedish Research Council SEE-6GIA 2024-06482.}
\thanks{Yanwen Chen and Stefano Maci are with the Department of Information
Engineering and Mathematics, University of Siena, 53100 Siena, Italy (e-mail: \mbox{yanwen.chen@unisi.it}, \mbox{macis@dii.unisi.it}). }
\thanks{Mats Gustafsson is with the Department of Electrical and Information Technology, Lund University, 22100 Lund, Sweden (e-mail: \mbox{mats.gustafsson@eit.lth.se}).}}

\maketitle

\begin{abstract}

The number of spatial degrees of freedom (NDoF) of electromagnetic communication channels in reflecting environments is investigated using a full-vector formulation based on dyadic Green's functions and equivalent electric and magnetic surface currents. The channel correlation eigenproblem characterizes the electromagnetic communication modes and their coupling strengths, while the leading-order NDoF is estimated geometrically from the receiver-accessible phase space volume in the electrically large regime. By combining two geometrical projection Jacobians, the
phase space volume is related to the mutual shadow area, establishing the equivalence between the two geometrical
NDoF formulations and yielding a two-polarization estimate
equal to twice the mutual shadow area divided by the
squared wavelength. The free-space mutual shadow formulation is extended to perfect electric conductor (PEC) reflecting environments through image construction and phase space ray tracing, accounting for specular reflections and geometrical visibility constraints. The formulation is validated against the eigenvalue spectra of full-vector electromagnetic channels involving planar PEC reflectors, a parabolic reflector, a parallel-plate waveguide, and a non-line-of-sight L-shaped corridor. The geometrical predictions agree with the observed spectral transitions across configurations involving multiple reflections, focusing, and blockage.

\end{abstract}


\section{Introduction}
\label{sec:introduction}

\IEEEPARstart{T}{he} number of degrees of freedom (NDoF)
characterizes the number of independent spatial modes available for electromagnetic communication between transmitting and receiving regions \cite{Miller00,PiestunMiller2000,Miller_19,Bucci89,Bucci_Migliore25,kaushik2026Book, Kildal_Maci_Martini,Maci_Martini_RS,Dardari20,Puggelli_DoF}. It provides a fundamental measure of the spatial multiplexing capability of multiple-input multiple-output (MIMO) systems, constrained by the physical dimensions of the antennas and the propagation environment~\cite{PoonBrodersenTse2005}. The communication modes are characterized by the
eigenvalue problem of the electromagnetic channel correlation operator, whose eigenvalues quantify
their coupling strengths. For electrically large configurations, the eigenvalue spectrum typically exhibits a transition from significant to rapidly decaying eigenvalues, whose location
provides the leading-order spatial NDoF \cite{GustafssonBrick2026}. An angular spectral interpretation of the channel eigenvalue distribution has also been investigated for line of sight (LoS) communications~\cite{TJCui2026}.
Although this mode count can be determined numerically through singular value decomposition (SVD)
of the channel operator, geometrical estimates provide a more direct physical interpretation of its dependence on the propagation environment.

The mathematical foundation for such estimates is closely related to the classical spatial--spectral
concentration problem developed by Slepian, Pollak, and Landau \cite{SlepianPollak_1,LandauPollak_3,Slepian_4}. Their theory characterizes the number of independent eigenfunctions that can be simultaneously concentrated within prescribed spatial and spectral regions, with their number determined by the corresponding spatial--spectral volume. This principle has also been extended to the transport of electromagnetic information, including multiple-scattering environments \cite{Franceschetti2015Laundau, kaushik2026Book}.
A complementary asymptotic estimate follows from Weyl's law, which relates the number of propagating
modes to the electrical dimensions of a region \cite{weyl1911asymptotische,Gustafsson2025Radiating}. These studies establish the connection between electromagnetic mode counts and the underlying spatial and spectral constraints.

The spatial--spectral connection can be described in phase space, where each propagation state on an aperture is characterized by its spatial position and tangential wavevector. In the geometrical-optics (GO) regime, the phase of a wavefield determines ray trajectories, along which localized wavefield components propagate in phase space~\cite{Arnold02}. Free-space propagation and ideal specular reflection induce canonical transformations that preserve the phase space measure \cite{Arnold02,Migliore2026Symplectic}. Phase space methods have also been applied to electromagnetic information transport and near-field communication, connecting geometrical propagation with the spectral properties of communication operators \cite{Gradoni_PRA,Gradoni_AP,Gradoni_SNR}. In this interpretation, the spatial NDoF is associated with the accessible phase space support 
between the transmitting and receiving regions.

The mutual shadow area provides a direct geometrical interpretation of the accessible phase space volume~\cite{Gustafsson2025Radiating,GustafssonBrick2026}. For communication between two regions in free space, under plane-wave illumination from a given direction, the transmitting and receiving regions cast shadows onto a common plane perpendicular to the propagation
direction. The overlap of these shadows, integrated over all directions, provides an asymptotic estimate of the spatial NDoF \cite{Gustafsson2025Shadow}. This formulation generalizes conventional paraxial estimates and applies to more general transmitting and receiving geometries. Reflecting environments introduce additional propagation paths that can redirect the transmitted phase space support toward the receiver. At the same time, geometrical blockage may restrict the physically accessible paths. Full-wave numerical studies of two-dimensional multiple-scattering environments have shown that scattering can modify the electromagnetic channel eigenvalue spectrum and increase the effective spatial DoF~\cite{XuJanaswamy2006}, motivating a geometrical description of the spatial NDoF in more general
propagation configurations.

An important consideration in extending these geometrical estimates is the vector nature of electromagnetic propagation. Although scalar Green's function models provide a convenient description of spatial coupling, general electromagnetic channel formulations retain the vector field interactions \cite{PiestunMiller2000,Miller00,Dardari20 ,GustafssonBrick2026}. In the present work, we employ full dyadic Green's functions and equivalent electric and magnetic surface currents to construct electromagnetic channels in
reflecting environments and validate the geometrical NDoF predictions against their eigenvalue spectra.

Specifically, this work extends the free-space mutual shadow formulation to reflecting environments and
establishes its equivalence to phase space mode counting for estimating the leading-order spatial NDoF between electrically large apertures. The accessible phase space support is identified by geometrically admissible ray trajectories satisfying the reflection and visibility constraints of the propagation environment. Through two geometrical
projection Jacobians, the phase space volume is related to the mutual shadow area, providing a direct geometrical estimate of the spatial NDoF accounting for both polarizations.

The formulation is validated against the eigenvalue spectra of electromagnetic channels in four reflecting environments
involving perfect electric conductor (PEC) surfaces: planar reflectors, a parabolic reflector, a parallel-plate waveguide, and a non-line-of-sight (NLoS) L-shaped parallel-plate corridor. These configurations encompass reflection and focusing by curved surfaces, infinitely many images of the transmitter, and NLoS propagation subject to geometrical blockage. The accessible phase space volume is evaluated through mutual shadow integration or phase space ray tracing and compared with the spectral transitions of the corresponding vector electromagnetic channels. These results demonstrate that the proposed formulation successfully predicts the leading-order spatial NDoF across the considered reflecting environments.

The remainder of this paper is organized as follows. Section~\ref{sec:electromagnetic_channel} establishes the full-vector electromagnetic channel formulation, including the correlation eigenproblem and the exact spectral sum rule. 
Section~\ref{sec:phase_space_mutual_shadow} develops the leading-order geometrical NDoF formulation based on phase space and mutual shadow area, including the paraxial approximation.
Section~\ref{sec:geometrical_evaluation} extends the formulation to reflecting environments through phase space ray tracing and image construction.
Section~\ref{sec:numerical_validation} validates the resulting geometrical NDoF estimates against electromagnetic channel eigenvalue spectra in four PEC reflecting environments. Finally, Section~\ref{sec:conclusion} summarizes the main findings.

\section{Electromagnetic Channel and Spatial Modes}
\label{sec:electromagnetic_channel}
The classical communication problem concerns electromagnetic propagation between two bounded transmitting and receiving regions.
The number of spatial degrees of freedom (NDoF) is defined here as the leading number of radiative spatial modes supported by the electromagnetic channel between the two regions, including both
transverse polarizations. This problem is particularly relevant to electrically large near-field communication systems, where the spatial variation of the propagation geometry generally hinders the use of conventional far-field approximations
\cite{Gradoni_PRA,Gustafsson2025Shadow}.

This section establishes the electromagnetic modal problem through the surface-equivalent channel formulation, followed by the correlation eigenproblem and the corresponding spectral sum rule.

\subsection{Surface-Equivalent Electromagnetic Channel}
\label{subsec:channel_operator}

In the general volume-to-volume formulation, the electromagnetic coupling between transmitting region $\varOmega_{\mathrm T}$ and receiving region $\varOmega_{\mathrm R}$ is described by a linear channel operator that maps electric and magnetic current densities supported in volume $\varOmega_{\mathrm T}$ to the resulting electric and magnetic fields in volume $\varOmega_{\mathrm R}$ \cite{GustafssonBrick2026}. This operator characterizes all possible electromagnetic field configurations achievable in this communication channel.

By the surface equivalence theorem, the electromagnetic sources within $\varOmega_{\mathrm T}$ can be replaced by equivalent electric and magnetic surface currents on its enclosing boundary $\partial\varOmega_{\mathrm T}$, which are determined by the tangential magnetic and electric fields respectively, such that the same exterior electromagnetic field is reproduced \cite{balanis2012advanced}. On the receiving side, assuming that $\varOmega_{\mathrm R}$ is source-free, the electromagnetic field within this region is \emph{uniquely} determined by its tangential electric and magnetic fields on the enclosing boundary $\partial\varOmega_{\mathrm R}$. The interior field can therefore be reconstructed from the equivalent surface currents associated with these boundary fields.

Consequently, by the surface equivalence theorem and the uniqueness of solutions to Maxwell's equations, a complete equivalent surface representation preserves the physical solution space of the original volume-to-volume communication problem. This motivates a surface formulation of the electromagnetic channel between the transmitting and receiving regions.

Accordingly, we represent the transmitting and receiving
regions by their enclosing surfaces $S_{\mathrm T}=\partial\varOmega_{\mathrm T}$ and $S_{\mathrm R}=\partial\varOmega_{\mathrm R}$, respectively. The electromagnetic coupling is formulated in terms
of equivalent electric and magnetic surface currents $\mathbf J(\mathbf r_{\mathrm T})$ and
$\mathbf M(\mathbf r_{\mathrm T})$ supported on $S_{\mathrm T}$, which generate electric and magnetic
fields on $S_{\mathrm R}$. Although these fields are related by Maxwell's equations
and do not represent independent sets of electromagnetic DoF, both are retained to provide a complete
surface-field representation of the channel.

We adopt an impedance normalization by scaling the electric surface currents and magnetic fields
by the free-space impedance $\eta_0$, placing the electric and magnetic components of both
source and observation vectors on the same dimensional scale. The electromagnetic channel is then expressed as
\begin{equation}
\begin{bmatrix}
\mathbf E_{\mathrm t}(\mathbf r_{\mathrm R})\\
\eta_0\mathbf H_{\mathrm t}(\mathbf r_{\mathrm R})
\end{bmatrix}
=
\int_{S_{\mathrm T}}
\mathbf G(\mathbf r_{\mathrm R},\mathbf r_{\mathrm T})
\cdot
\begin{bmatrix}
\eta_0\mathbf J(\mathbf r_{\mathrm T})\\
\mathbf M(\mathbf r_{\mathrm T})
\end{bmatrix}
\,\mathrm dS_{\mathrm T},
\label{eq:channel_operator}
\end{equation}
where $\mathbf E_{\mathrm t}$ and $\mathbf H_{\mathrm t}$ are the tangential electric and magnetic field components on the receiving surface $S_{\mathrm R}$. The combined impedance-normalized Green's function is
\begin{equation}
\mathbf G=
\begin{bmatrix}
\mathbf G_{\mathrm t}^{\mathrm{EJ}} &
\mathbf G_{\mathrm t}^{\mathrm{EM}}\\
\mathbf G_{\mathrm t}^{\mathrm{HJ}} &
\mathbf G_{\mathrm t}^{\mathrm{HM}}
\end{bmatrix},
\label{eq:G}
\end{equation}
where the superscripts identify the observed field and source types, and the dyadic blocks are projected onto the corresponding surface tangential directions.
The normalized free-space Green's functions are given in Appendix~\ref{app:dyadic_green}.
The resulting linear mapping between the normalized surface currents and fields defines the channel
operator $\mathcal H$ considered in this work.

\subsection{Correlation Eigenproblem and Spectral Sum Rule}
\label{subsec:correlation_eigenproblem}

The singular value decomposition (SVD) of $\mathcal H$ characterizes the spatial communication modes between the transmitting and receiving surfaces. The transmit-side modes are its right singular functions, or equivalently, the eigenfunctions of the correlation operator $\mathcal H^\dagger\mathcal H$, whose spatial kernel is
\begin{equation}
\mathbf R(\mathbf r_{\mathrm T,2},\mathbf r_{\mathrm T,1})
=
\int_{S_{\mathrm R}}
\mathbf G^\dagger(\mathbf r_{\mathrm R},\mathbf r_{\mathrm T,2})
\cdot
\mathbf G(\mathbf r_{\mathrm R},\mathbf r_{\mathrm T,1})
\,\mathrm dS_{\mathrm R}.
\label{eq:correlation_operator}
\end{equation}
Here, the superscript $\dagger$ indicates the adjoint with respect to the surface inner products.
The kernel $\mathbf R$ characterizes the correlation between the combined electric and magnetic field
responses to surface-current excitations at two transmitting points. The corresponding eigenvalue
problem is
\begin{equation}
\int_{S_{\mathrm T}}
\mathbf R(\mathbf r_{\mathrm T,2},\mathbf r_{\mathrm T,1})
\cdot
\begin{bmatrix}
\eta_0\mathbf J_n(\mathbf r_{\mathrm T,1})\\
\mathbf M_n(\mathbf r_{\mathrm T,1})
\end{bmatrix}
\,\mathrm dS_{\mathrm T,1}
=
\zeta_n
\begin{bmatrix}
\eta_0\mathbf J_n(\mathbf r_{\mathrm T,2})\\
\mathbf M_n(\mathbf r_{\mathrm T,2})
\end{bmatrix},
\label{eq:correlation_eigenproblem}
\end{equation}
where $\mathbf J_n$ and $\mathbf M_n$ are the electric and magnetic surface-current distributions of the $n$th transmit-side mode, and $\zeta_n$ is the corresponding dimensionless eigenvalue of
$\mathcal H^\dagger\mathcal H$, characterizing its coupling strength. For electrically large surfaces, the eigenvalue
spectrum typically contains a set of significant modes followed by a rapid decay, with the location
of the transition providing the leading-order spatial NDoF~\cite{GustafssonBrick2026}.

For spatially separated transmitting and receiving
surfaces, the eigenvalues $\zeta_n$ of
$\mathcal H^\dagger\mathcal H$ satisfy the spectral
sum rule
\begin{equation}
    \sum_n\zeta_n
    =\operatorname{Tr}(\mathcal H^\dagger\mathcal H)
    =\|\mathcal H\|_{\mathrm{HS}}^2
    =\int_{S_{\mathrm T}}\!\int_{S_{\mathrm R}}
    \|\mathbf G(\mathbf r_{\mathrm R},
    \mathbf r_{\mathrm T})\|_{\mathrm F}^2
    \,\mathrm dS_{\mathrm R}\,\mathrm dS_{\mathrm T}.
    \label{eq:general_sum_rule}
\end{equation}
Here, $\mathbf G$ is given by \eqref{eq:G}, while
$\|\cdot\|_{\mathrm{HS}}$ and $\|\cdot\|_{\mathrm F}$
are the Hilbert--Schmidt norm of an operator and
the Frobenius norm of a matrix, respectively.
This basis-independent identity extends the scalar
spectral sum rule to the vector electromagnetic
channel~\cite{Miller00,GustafssonBrick2026, Dardari20,
KuangMillerMiller2025}.

\section{Geometrical Formulation of Spatial DoF}
\label{sec:phase_space_mutual_shadow}

For electrically large configurations, wave propagation admits a semiclassical asymptotic description in which the field is characterized by a rapidly varying phase modulated by a comparatively slowly varying amplitude. The phase satisfies the eikonal equation, while the subsequent transport equation determines the leading amplitude \cite{born1999principles}. The characteristics of the eikonal equation define the geometrical-optics (GO) ray trajectories, providing the corresponding Hamiltonian description in terms of position and local wavevector \cite{born1999principles,Migliore2026Symplectic}. This asymptotic geometrical description provides the basis for the leading-order phase space mode count developed below.

This section develops a geometrical formulation for estimating the leading-order spatial NDoF of electromagnetic communication channels. The accessible phase space support is first
constructed from the local tangential wavevector and receiver-accessible spectrum.
The resulting phase space volume is then related to the mutual shadow area through two geometrical projection Jacobians, followed by its paraxial limit.

\subsection{Local Tangential Wavevector and Accessible Spectrum}
\label{subsec:accessible_spectrum}

\begin{figure}
    \centering
    \includegraphics[width=1\linewidth]{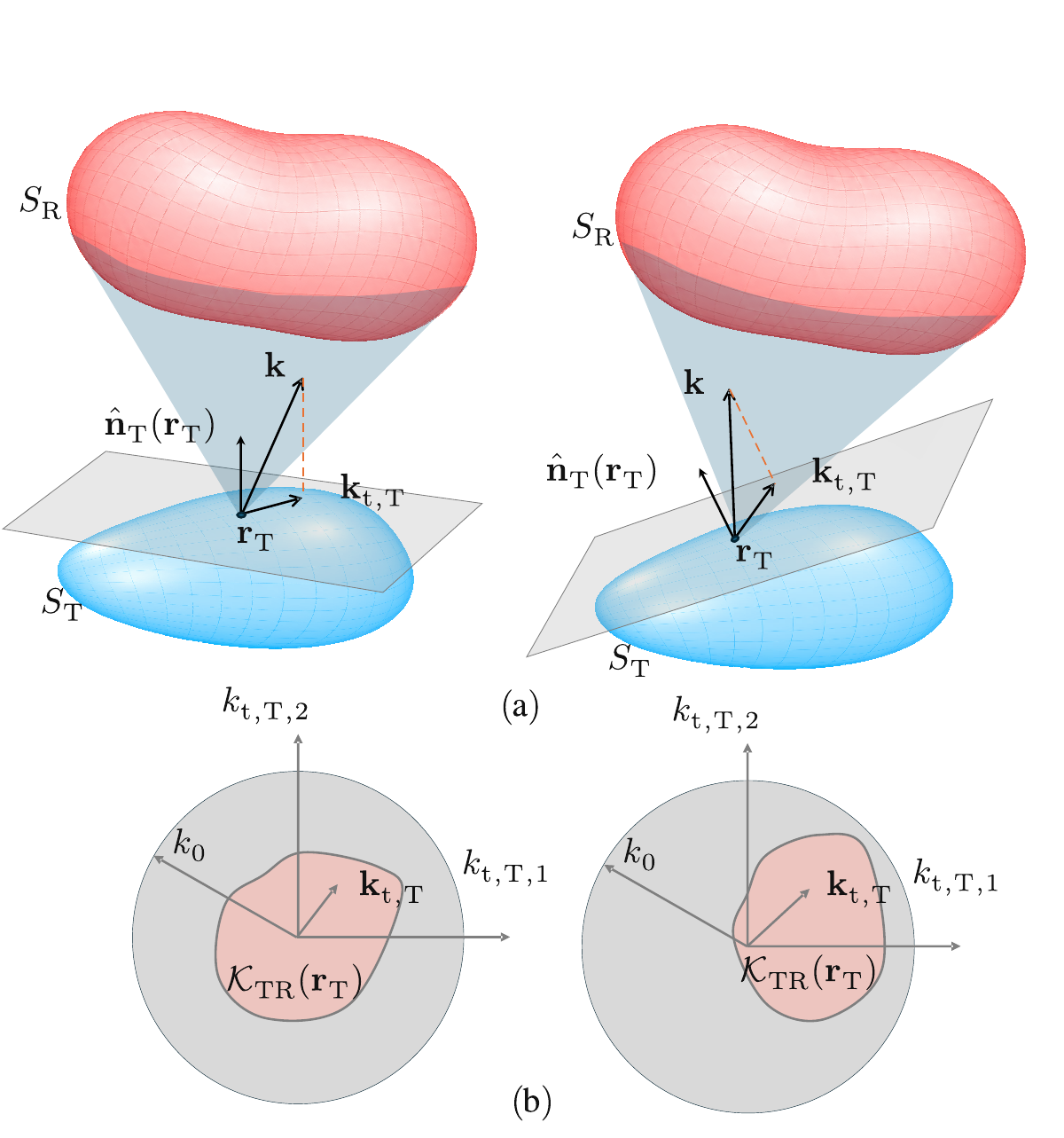}
    \caption{Spatial and spectral representations of the electromagnetic
channel. (a) Transmitting and receiving surfaces
$S_{\mathrm T}$ and $S_{\mathrm R}$, with the wavevector $\mathbf k$ at $\mathbf r\in S_{\mathrm T}$ and its tangential projection
$\mathbf k_{\mathrm t}$ onto the local tangent plane. (b)
Tangential wavevector plane showing the visible region
$|\mathbf k_{\mathrm t}|\leq k_0$ and the channel-accessible subset
$\mathcal K_{\mathrm{TR}}$; the indicated $\mathbf k_{\mathrm t}$ corresponds to the
tangential component in (a).}
    \label{fig:Spatial_Spectral}
\end{figure}

Assuming a smooth transmitting surface $S_{\mathrm T}$, we introduce the local tangent plane at $\mathbf r_{\mathrm T}\in S_{\mathrm T}$
as the plane passing through $\mathbf r_{\mathrm T}$ and perpendicular to the outward unit normal $\hat{\mathbf n}_{\mathrm T}(\mathbf r_{\mathrm T})$.
The propagation wavevector $\mathbf k$ at this point can then be decomposed into its tangential projection $\mathbf k_{\mathrm{t,T}}$ onto this plane
and its normal component $k_{\mathrm{n,T}}$ as
\begin{equation}
    \mathbf k
    =
    \mathbf k_{\mathrm {t,T}}(\mathbf r_{\mathrm T})
    +
    k_{\mathrm {n,T}}(\mathbf r_{\mathrm T})
    \hat{\mathbf n}_{\mathrm T}(\mathbf r_{\mathrm T}).
    \label{eq:local_wavevector_decomposition}
\end{equation}
Accordingly, the pair $(\mathbf r_{\mathrm T},\mathbf k_{\mathrm {t,T}})$ identifies a local
propagating state through its surface position and propagation direction, as illustrated in Fig.~\ref{fig:Spatial_Spectral}(a). For a propagating component in free space, the outgoing normal wavenumber at $\mathbf r_{\mathrm T}$ is $k_{\mathrm {n,T}} =\sqrt{k_0^2-|\mathbf k_{\mathrm {t,T}}|^2}$. The propagation condition requires $k_{\mathrm {n,T}}$ to be real and
therefore naturally defines a disk of radius $k_0$ in tangential wavevector space, commonly referred to as the visible region as $\mathcal K_{\mathrm{vis},\mathrm T}(\mathbf r_{\mathrm T})
=\{\mathbf k_{\mathrm {t,T}}:|\mathbf k_{\mathrm {t,T}}|\leq k_0\}$. The same construction applies to a homogeneous, isotropic, lossless background medium by replacing $k_0$ with the corresponding background wavenumber.

The visible region defined above characterizes the spectrum locally available for propagation from $S_{\mathrm T}$. The channel operator $\mathcal H$ defined in \eqref{eq:channel_operator}, however, includes both propagating and evanescent field contributions. To establish the leading-order radiative spatial NDoF of this channel problem, we consider the regime in which propagating components dominate the mode count, generally outside the reactive near-field region. In this regime, evanescent fields remain part of the full channel, but their influence is comparatively weaker. Accordingly, we retain only the propagating spectral components of the Green's functions for the leading-order mode count, while the original channel operator $\mathcal H$ remains intact.

For communication between $S_{\mathrm T}$ and $S_{\mathrm R}$, the propagating spectrum is further restricted by the receiving surface $S_{\mathrm R}$ and the corresponding propagation environment, as represented by the corresponding Green's functions. Only the propagating spectral components that couple $S_{\mathrm T}$ to $S_{\mathrm R}$ contribute to the leading-order radiative mode count. These propagating contributions generally occupy only a subset of the visible region, as illustrated in Fig.~\ref{fig:Spatial_Spectral}(b). We denote this receiver-accessible tangential wavevector region at $\mathbf r_{\mathrm T}$ by $\mathcal K_{\mathrm{TR}}(\mathbf r_{\mathrm T})$, with $\mathcal K_{\mathrm{TR}}(\mathbf r_{\mathrm T}) \subseteq \mathcal K_{\mathrm{vis},\mathrm T}(\mathbf r_{\mathrm T})$.

The resulting problem is therefore subject to simultaneous spatial and spectral restrictions. The spatial support is confined to the transmitting surface $S_{\mathrm T}$, while the accessible spectral support is described by $\mathcal K_{\mathrm{TR}}(\mathbf r_{\mathrm T})$. The corresponding spatial NDoF can be estimated
asymptotically from the volume of the accessible phase space, as developed in the following subsection. 

\subsection{Generalized Phase Space Formulation}
\label{subsec:Generalized_Phase_Space}

The joint domain of the transmitting position
$\mathbf r_{\mathrm T}$ and its local tangential wavevector $\mathbf k_{\mathrm{t,T}}$ is referred to as the transmit-side \emph{phase space}
\cite{Migliore2026Symplectic,Gradoni_PRA,Gradoni_SNR}. 

The classical Slepian--Landau concentration problem,
as reviewed in Section~S1
of the Supplementary Material, considers a planar transmitting surface whose spectral support is independent of spatial position, such that $\mathcal K_{\mathrm{TR}}(\mathbf r_{\mathrm T})
=\mathcal K_{\mathrm{TR}}$.
The phase space support then forms the Cartesian
product
$\Gamma_{\mathrm{TR}}
=S_{\mathrm T}\times\mathcal K_{\mathrm{TR}}$,
with volume
$V_{\mathrm{PS}}
=|S_{\mathrm T}|\,|\mathcal K_{\mathrm{TR}}|$.
For a scalar field, the asymptotic mode count
is rigorously established through the concentration
eigenvalue distribution.

For a planar transmitting surface in a general
propagation configuration, however, the receiver-accessible
tangential wavevector region
$\mathcal K_{\mathrm{TR}}(\mathbf r_{\mathrm T})$
generally varies with the transmitting position
$\mathbf r_{\mathrm T}\in S_{\mathrm T}$.
The corresponding transmitting-side phase space support
is defined as
$\Gamma_{\mathrm{TR}}
=\{(\mathbf r_{\mathrm T},\mathbf k_{\mathrm{t,T}}):
\mathbf r_{\mathrm T}\in S_{\mathrm T},\,
\mathbf k_{\mathrm{t,T}}\in
\mathcal K_{\mathrm{TR}}(\mathbf r_{\mathrm T})\}$.
Unlike the classical separable case, this region
generally does not form a Cartesian product of
spatial and spectral supports.
Its phase space volume \cite{Gradoni_SNR,Migliore2026Symplectic} is
\begin{equation}
    V_{\mathrm{PS}}
    =
    \int_{S_{\mathrm T}}
    \int_{\mathcal K_{\mathrm{TR}}(\mathbf r_{\mathrm T})}
    \mathrm d\mathbf k_{\mathrm{t,T}}\,
    \mathrm dS_{\mathrm T}.
    \label{eq:general_phase_space_volume}
\end{equation}

In general propagation geometries, the Green's function operator does not necessarily reduce to the Fourier restriction operator used in the
classical Slepian--Landau concentration
problem~\cite{Gradoni_PRA,Gradoni_AP}.
In the electrically large regime, however, propagation through a lossless medium, including specular reflections from ideal PEC surfaces, induces canonical mappings between the transmit- and receive-side phase spaces \cite{Gradoni_AP,Migliore2026Symplectic}.
Mapping the receiver surface restriction back through these transformations yields the generally non-separable
transmit-side phase space support $\Gamma_{\mathrm{TR}}$ \cite{Gradoni_PRA}. For the one-dimensional scalar configuration studied
in~\cite{Gradoni_PRA}, the resulting phase space region can be rectified by a canonical change of coordinates
into the rectangular form of the classical Slepian--Landau problem. The transformation preserves the phase space volume
\cite{Gradoni_PRA,Migliore2026Symplectic}, leaving the leading-order mode count unchanged.

The phase space description also extends locally to general smooth closed surfaces in the electrically large regime. Provided that the surface geometry varies sufficiently slowly on the local wavelength scale
\cite{Migliore2026Symplectic}, the tangential wavevector $\mathbf k_{\mathrm{t,T}}$ and the accessible region $\mathcal K_{\mathrm{TR}}(\mathbf r_{\mathrm T})$ are defined with respect to the local tangent plane at each $\mathbf r_{\mathrm T}\in S_{\mathrm T}$. The phase space volume in \eqref{eq:general_phase_space_volume} therefore remains applicable.

In the electrically large limit, Weyl's phase space
counting principle associates each scalar mode with
an elementary phase space volume of $(2\pi)^2$
\cite{weyl1911asymptotische,Gustafsson2025Shadow,
Migliore2026Symplectic,Hormander1985}.
The leading-order mode count \emph{per polarization} is therefore given by $V_{\mathrm{PS}}/(2\pi)^2$,
with lower order asymptotic corrections.
For the general electromagnetic channel, we adopt the corresponding leading-order NDoF, including both transverse polarizations, as
\begin{equation}
    N_{\mathrm{DoF}}
    \simeq
    \frac{2V_{\mathrm{PS}}}{(2\pi)^2}
    =
    \frac{2}{(2\pi)^2}
    \int_{S_{\mathrm T}}
    \int_{\mathcal K_{\mathrm{TR}}(\mathbf r_{\mathrm T})}
    \mathrm d\mathbf k_{\mathrm{t,T}}\,
    \mathrm dS_{\mathrm T}.
    \label{eq:phase_space_volume_DoF}
\end{equation}

\subsection{Mutual Shadow Area as a Geometrical Interpretation of Phase Space Volume}
\label{subsec:shadow_mutual_shadow}

The phase space formulation established above characterizes the transmit--receive coupling through the accessible support $\Gamma_{\mathrm{TR}}$ and its volume $V_{\mathrm{PS}}$. This phase space volume has a direct geometrical interpretation through the mutual shadow area of the transmitting and receiving surfaces
\cite{Gustafsson2025Radiating,Gustafsson2025Shadow,GustafssonBrick2026}.

\begin{figure}
    \centering
    \includegraphics[width=1\linewidth]{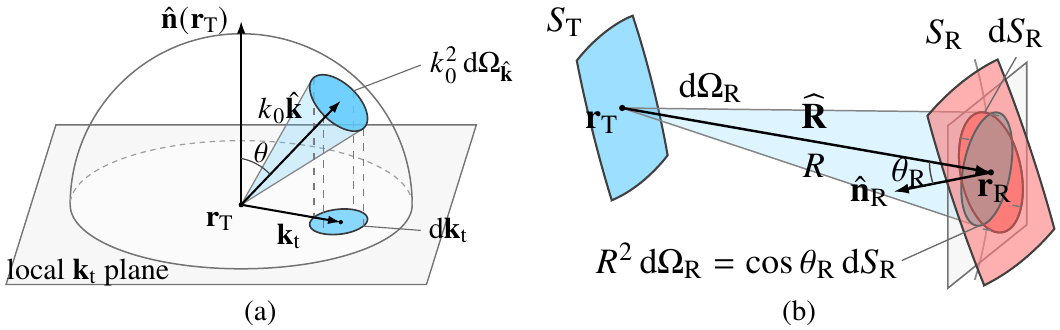}
    \caption{Two geometrical projections underlying the mutual shadow
    representation of phase space.
    (a) At $\mathbf r_{\mathrm T}\in S_{\mathrm T}$, the wavevector
    $k_0\hat{\mathbf k}$ projects onto the local tangential wavevector
    plane as $\mathbf k_{\mathrm {t,T}}$, mapping the $k$-sphere element
    $k_0^2\,\mathrm d\Omega_{\hat{\mathbf k}}$ to
    $\mathrm d\mathbf k_{\mathrm {t,T}}$ through the transmitter
    projection factor.
    (b) For $\mathbf r_{\mathrm T}\in S_{\mathrm T}$ and
    $\mathbf r_{\mathrm R}\in S_{\mathrm R}$, the receiver element
    $\mathrm dS_{\mathrm R}$ projects onto the plane normal to
    $\widehat{\mathbf R}$, giving the apparent area
    $\bigl|\hat{\mathbf n}_{\mathrm R}\cdot\widehat{\mathbf R}\bigr|
    \mathrm dS_{\mathrm R}$ and the corresponding spherical area
    $R^2\,\mathrm d\Omega_{\mathrm R}$ subtended at
    $\mathbf r_{\mathrm T}$.}
    \label{fig:shadow_mutual_shadow}
\end{figure}

Consider the local spectral projection at a point
$\mathbf r_{\mathrm T}$ on $S_{\mathrm T}$.
A propagating wavevector $\mathbf k=k_0\hat{\mathbf k}$
is projected onto the local tangent plane to obtain
the tangential wavevector $\mathbf k_{\mathrm{t,T}}$.
With the outward-oriented surface normal
$\hat{\mathbf n}_{\mathrm T}$, the corresponding
Jacobian is
\begin{equation}
    \mathrm d\mathbf k_{\mathrm{t,T}}
    =
    k_0^2
    [\hat{\mathbf n}_{\mathrm T}\cdot\hat{\mathbf k}]_+
    \,\mathrm d\Omega_{\hat{\mathbf k}},
    \label{eq:solid_angle_kt_jacobian}
\end{equation}
where $[x]_+=\max(x,0)$ selects outward propagation
directions. Defining $\theta_{\mathrm T}$ as the angle
between $\hat{\mathbf n}_{\mathrm T}$ and
$\hat{\mathbf k}$, the projection factor reduces to
$\cos\theta_{\mathrm T}$ on the outward hemisphere.
As illustrated in Fig.~\ref{fig:shadow_mutual_shadow}(a),
\eqref{eq:solid_angle_kt_jacobian} maps the differential
$k$-sphere area $k_0^2\,\mathrm d\Omega_{\hat{\mathbf k}}$ onto the local tangential wavevector area
$\mathrm d\mathbf k_{\mathrm{t,T}}$ through the
projected-area factor $[\cos\theta_{\mathrm T}]_+$.
Integration of the local projection over all accessible
outward directions gives the shadow area representation
of the phase space volume associated with an isolated
transmitting surface \cite{Gustafsson2025Shadow,Migliore2026Symplectic}.

For the general communication problem, only directions
connecting the transmitting and receiving surfaces
contribute to the channel. Consider two points
$\mathbf r_{\mathrm T}\in S_{\mathrm T}$ and
$\mathbf r_{\mathrm R}\in S_{\mathrm R}$, and define
$\mathbf R=\mathbf r_{\mathrm R}-\mathbf r_{\mathrm T}$,
$R=|\mathbf R|$, and $\widehat{\mathbf R}=\mathbf R/R$.
For a fixed transmitting point $\mathbf r_{\mathrm T}$, accounting for its inward-facing orientation, the solid angle subtended by a receiver surface element
$\mathrm dS_{\mathrm R}$ is
\begin{equation}
    \mathrm d\Omega_{\mathrm R}
    =
    \frac{
    [-\hat{\mathbf n}_{\mathrm R}\cdot\widehat{\mathbf R}]_+
    }{R^2}
    \,\mathrm dS_{\mathrm R}.
    \label{eq:rx_solid_angle}
\end{equation}
Define $\theta_{\mathrm R}$ as the angle between the inward receiver normal $-\hat{\mathbf n}_{\mathrm R}$ and the propagation direction $\widehat{\mathbf R}$. As illustrated in Fig.~\ref{fig:shadow_mutual_shadow}(b), the receiver element $\mathrm dS_{\mathrm R}$ is projected onto the plane normal to $\widehat{\mathbf R}$, giving the
projected area $[\cos\theta_{\mathrm R}]_+ \,\mathrm dS_{\mathrm R}$. This equals the differential area $R^2\,\mathrm d\Omega_{\mathrm R}$ on the sphere of radius $R$ centered at $\mathbf r_{\mathrm T}$. Thus, $[\cos\theta_{\mathrm R}]_+$ is the receiver-side area projection factor, while $1/R^2$ converts the
projected area into solid angle. Locally, this transverse-plane projection resembles the paraxial formulation, but is exact for arbitrary incidence angles. Equation~\eqref{eq:rx_solid_angle} therefore provides the first projection Jacobian, mapping the receiver surface element $\mathrm dS_{\mathrm R}$ to its subtended solid angle $\mathrm d\Omega_{\mathrm R}$.

The resulting solid-angle element is then mapped
onto the transmit-side tangential wavevector plane
using \eqref{eq:solid_angle_kt_jacobian}, with
$\hat{\mathbf k}=\widehat{\mathbf R}$, giving the
second projection Jacobian as
\begin{equation}
    \mathrm d\mathbf k_{\mathrm{t,T}}
    =
    k_0^2
    [\hat{\mathbf n}_{\mathrm T}\cdot\widehat{\mathbf R}]_+
    \,\mathrm d\Omega_{\mathrm R}.
    \label{eq:tx_spectral_projection}
\end{equation}
Combining the two projection Jacobians yields
\begin{equation}
    \mathrm d\mathbf k_{\mathrm{t,T}}
    =
    k_0^2
    \frac{
    [\hat{\mathbf n}_{\mathrm T}\cdot\widehat{\mathbf R}]_+
    [-\hat{\mathbf n}_{\mathrm R}\cdot\widehat{\mathbf R}]_+
    }{R^2}
    \,\mathrm dS_{\mathrm R}.
    \label{eq:TR_jacobian}
\end{equation}
The two projection factors in \eqref{eq:TR_jacobian} have distinct geometrical
origins. The receiver factor $ [-\hat{\mathbf n}_{\mathrm R}\cdot\widehat{\mathbf R}]_+$ maps the physical receiver element to its subtended solid angle, while the transmitter factor $[\hat{\mathbf n}_{\mathrm T}\cdot\widehat{\mathbf R}]_+$ maps that solid angle to tangential wavevector area at $\mathbf r_{\mathrm T}$. The positive-part notation $[\,\cdot\,]_+$ restricts the propagation along $\widehat{\mathbf R}$ to outward directions from $S_{\mathrm T}$ and inward directions into $S_{\mathrm R}$.

For two spatially separated convex regions bounded by $S_{\mathrm T}$ and $S_{\mathrm R}$, with no intervening obstructions, the mutual shadow area \cite{Gustafsson2025Shadow} is defined as
\begin{equation}
    A_{\mathrm{TR}}
    =
    \int_{S_{\mathrm T}}\!\int_{S_{\mathrm R}}
    \frac{
    [\hat{\mathbf n}_{\mathrm T}\cdot\widehat{\mathbf R}]_+
    [-\hat{\mathbf n}_{\mathrm R}\cdot\widehat{\mathbf R}]_+
    }{R^2}
    \,\mathrm dS_{\mathrm R}\,\mathrm dS_{\mathrm T}.
    \label{eq:TR_shadow}
\end{equation}
Substituting \eqref{eq:TR_jacobian} into \eqref{eq:general_phase_space_volume} and using \eqref{eq:TR_shadow} yields $V_{\mathrm{PS}}=k_0^2A_{\mathrm{TR}}$. Combining this relation with the phase space mode count in \eqref{eq:phase_space_volume_DoF} gives the leading-order electromagnetic NDoF estimate as
\begin{equation}
N_{\mathrm{DoF}}\simeq N_A
= {2A_{\mathrm{TR}}}/{\lambda^2}.
\label{eq:DoF_shadow}
\end{equation}
Here, $N_A$ is the geometrical NDoF estimate based on the mutual shadow area, including both transverse polarizations. For nonconvex regions, a visibility function $\chi_{\mathrm{vis}}$ must be included in the integrand of \eqref{eq:TR_shadow} to account for self-blockage and multiple surface intersections. It equals unity for unobstructed connections reaching the receiver at their first intersection and zero otherwise.

An equivalent directional representation of the mutual shadow area can be obtained by grouping the accessible propagation paths according to their direction $\hat{\mathbf k}$ \cite{Gustafsson2025Shadow,GustafssonBrick2026}.
Reversing the order of integration in \eqref{eq:general_phase_space_volume} and using the spectral projection Jacobian in
\eqref{eq:solid_angle_kt_jacobian} gives
\begin{equation}
V_{\mathrm{PS}}
=k_0^2\!\int_{4\pi}\!\int_{S_{\mathrm T}}
\chi_{\mathrm{TR}}(\mathbf r_{\mathrm T},\hat{\mathbf k})
[\hat{\mathbf n}_{\mathrm T}\cdot\hat{\mathbf k}]_+
\,\mathrm dS_{\mathrm T}\,\mathrm d\Omega_{\hat{\mathbf k}}.
\label{eq:phase_space_directional}
\end{equation}
Here, $\chi_{\mathrm{TR}}(\mathbf r_{\mathrm T},\hat{\mathbf k})$ equals unity for propagation directions accessible
from $\mathbf r_{\mathrm T}$ to $S_{\mathrm R}$, and zero otherwise. The angular integration extends over the full solid angle $4\pi$, while
the positive-part projection factor selects locally outward propagation directions.

\begin{figure}
    \centering
    \includegraphics[width=\linewidth]
{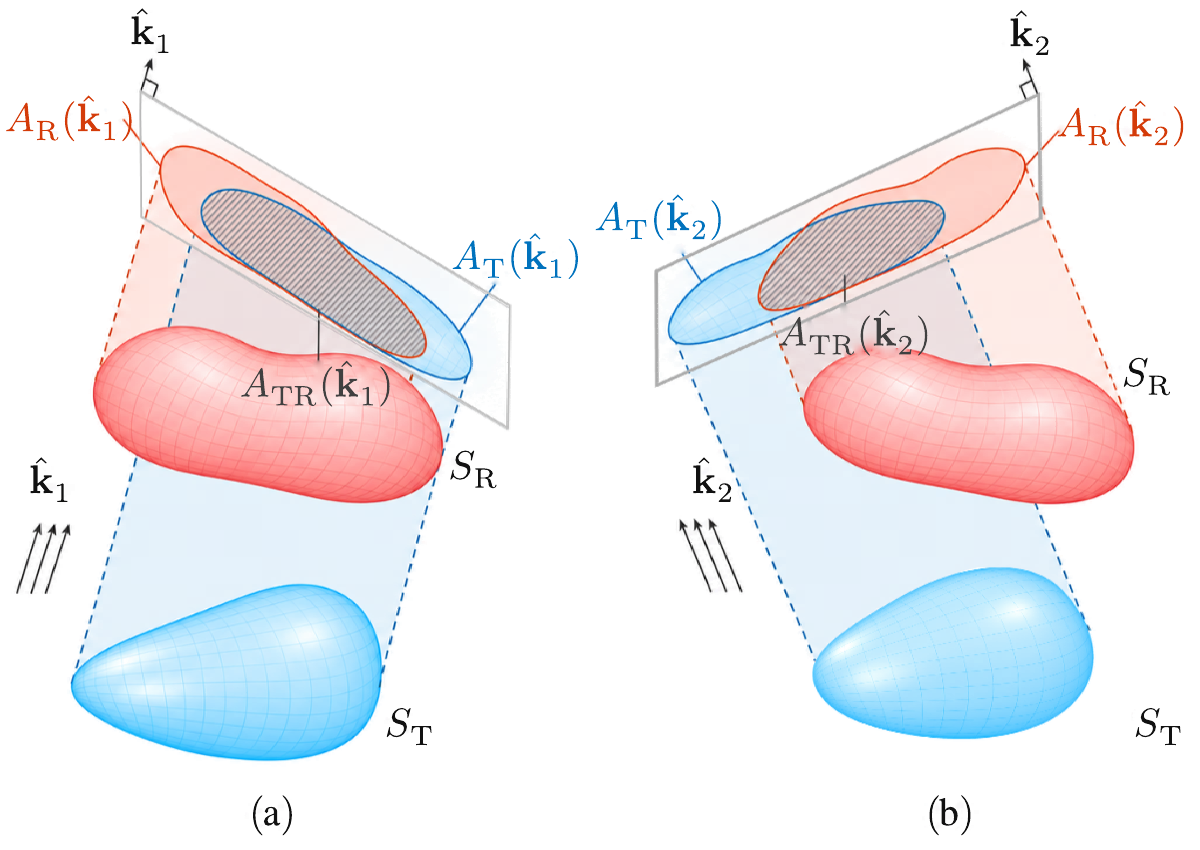}
    \caption{Directional mutual shadow area for two propagation directions $\hat{\mathbf k}_1$ and $\hat{\mathbf k}_2$. The transmitting and receiving regions are projected onto the plane normal to each direction. Their overlap, restricted to rays propagating from the transmitter to the receiver, defines $A_{\mathrm{TR}}(\hat{\mathbf k})$.}
    \label{fig:directional_mutual_shadow}
\end{figure}

For each fixed $\hat{\mathbf k}$, the inner spatial integral over $S_\mathrm T$ gives the projected transmitting area accessible to the receiver. For the convex regions considered above, this corresponds to the overlap of their projections onto the plane normal to $\hat{\mathbf k}$, restricted to propagation from the transmitter
to the receiver along $\hat{\mathbf k}$. Defining the projected regions as $\mathcal A_{\mathrm T}(\hat{\mathbf k})$ and
$\mathcal A_{\mathrm R}(\hat{\mathbf k})$, the directional mutual shadow area illustrated in Fig.~\ref{fig:directional_mutual_shadow} is
\begin{equation}
A_{\mathrm{TR}}(\hat{\mathbf k})
=
\operatorname{Area}\left[
\mathcal A_{\mathrm T}(\hat{\mathbf k})
\cap
\mathcal A_{\mathrm R}(\hat{\mathbf k})
\right],
\label{eq:directional_mutual_shadow}
\end{equation}
where only the overlap corresponding to propagation from the transmitter $S_{\mathrm T}$ to the receiver $S_{\mathrm R}$ is counted.

Integrating these directional overlaps over the full solid angle $4\pi$ recovers the mutual shadow area defined by the double surface integral in
\eqref{eq:TR_shadow}:
\begin{equation}
A_{\mathrm{TR}}
=\int_{4\pi}A_{\mathrm{TR}}(\hat{\mathbf k})
\,\mathrm d\Omega_{\hat{\mathbf k}}.
\label{eq:directional_shadow_integral}
\end{equation}
The surface-pair and directional formulations thus provide equivalent geometrical representations of the mutual shadow area governing the NDoF estimate in \eqref{eq:DoF_shadow}.

\subsection{Paraxial Approximation of the Mutual Shadow Area}
\label{subsec:paraxial}

The paraxial approximation applies when the propagation directions connecting the transmitting and receiving apertures are confined to a narrow angular range $\Delta\theta\ll1$ around a central propagation
direction $\widehat{\mathbf R}_0$. In this regime, the propagation direction and separation vary only weakly across the aperture pair and may be approximated by $\widehat{\mathbf R}\approx\widehat{\mathbf R}_0$ and $R\approx R_0$, respectively. In the mutual shadow construction, each differential receiver surface element $\mathrm dS_{\mathrm R}$ is projected onto the plane transverse to its
local propagation direction $\widehat{\mathbf R}$. Under the paraxial approximation, these local transverse planes are represented by the common plane normal to the central propagation direction $\widehat{\mathbf R}_0$. The surface normals $\hat{\mathbf n}_{\mathrm T}$ and $\hat{\mathbf n}_{\mathrm R}$ are constant for planar transmitting and receiving apertures. The mutual shadow kernel in \eqref{eq:TR_shadow} can then be taken outside the surface integrals, giving
\begin{equation}
    A_{\mathrm{TR}}
    \approx
    \frac{
    A_{\mathrm T}A_{\mathrm R}
    [\hat{\mathbf n}_{\mathrm T}\cdot\widehat{\mathbf R}_0]_+
    [-\hat{\mathbf n}_{\mathrm R}\cdot\widehat{\mathbf R}_0]_+
    }{R_0^2}.
    \label{eq:paraxial_mutual_shadow}
\end{equation}
Thus, the paraxial mutual shadow area is the product of the two projected aperture areas,
restricted to outward transmission and inward reception, divided by their squared central
separation $R_0^2$.
Paraxiality is determined by the angular spread about
$\widehat{\mathbf R}_0$, rather than by the orientation of $\widehat{\mathbf R}_0$ relative to the aperture normals. The central propagation direction may therefore be highly oblique while the ray
family remains paraxial.

For parallel, mutually facing transmitting and receiving apertures, the surface normals satisfy
$\hat{\mathbf n}_{\mathrm T}=-\hat{\mathbf n}_{\mathrm R}$. Defining $\theta_0$ as the angle between
$\hat{\mathbf n}_{\mathrm T}$ and the central propagation direction $\widehat{\mathbf R}_0$,
\eqref{eq:paraxial_mutual_shadow} reduces to
\begin{equation}
    A_{\mathrm{TR}}
    \approx
    \frac{A_{\mathrm T}A_{\mathrm R}}{R_0^2}
    \cos^2\theta_0 .
    \label{eq:paraxial_parallel_shadow}
\end{equation}
The broadside case $\theta_0=0$ recovers the standard paraxial expression $A_{\mathrm{TR}}\approx A_{\mathrm T}A_{\mathrm R}/R_0^2$, widely used for well-separated parallel apertures \cite{Miller00,Miller_19,Gustafsson2025Shadow}. The corresponding two-polarization NDoF estimate follows from \eqref{eq:DoF_shadow}.

\section{Geometrical Evaluation of Spatial DoF in Reflecting Environments}
\label{sec:geometrical_evaluation}

The preceding section established the geometrical
formulation of spatial NDoF in terms of the accessible
phase space volume and mutual shadow area in free space.
The same formulation can be extended to reflecting
environments, where reflections may enlarge the
accessible phase space and thereby increase the spatial
NDoF.
In this section, two complementary approaches are
presented: phase space ray tracing, which directly
evaluates the accessible phase space volume, and
mutual shadow evaluation through equivalent image
construction.

\subsection{Phase Space Ray Tracing}
\label{subsec:GO_interpretation}

The accessible phase space volume can be evaluated directly through geometrical ray tracing \cite{Gradoni_SNR}. In this paper, we consider only free-space propagation,
with rays following straight trajectories and undergoing specular reflections at perfect electric conductors (PECs).
For a geometrical path connecting
$\mathbf r_{\mathrm T}\in S_{\mathrm T}$ and
$\mathbf r_{\mathrm R}\in S_{\mathrm R}$,
the corresponding Green's function contribution
has the leading-order GO form
\begin{equation}
    \mathbf G(\mathbf r_{\mathrm R},\mathbf r_{\mathrm T})
    \simeq
    \mathbf A(\mathbf r_{\mathrm R},\mathbf r_{\mathrm T})
    \mathrm e^{-\mathrm j k_0
    \mathcal S(\mathbf r_{\mathrm R},\mathbf r_{\mathrm T})},
    \label{eq:GO_single_path}
\end{equation}
where $\mathcal S$ is the optical path length and
$\mathbf A$ is the corresponding dyadic amplitude
\cite{born1999principles}.
In the presence of multiple propagation paths,
the GO Green's function is approximated by the
superposition of their individual contributions.

Each ray is specified by its initial transmitting-side
phase space coordinates
$(\mathbf r_{\mathrm T},\mathbf k_{\mathrm{t,T}})$,
with the outward normal wavevector component determined
by the free-space dispersion relation $\mathbf{k}\!\cdot\!\mathbf{k}\!=\!k_0^2$.
The corresponding trajectory is then traced through
the propagation environment to determine whether it
reaches the receiving surface $S_{\mathrm R}$.

Both free-space propagation and specular reflection
are canonical transformations in ray phase space.
By Liouville's theorem, the phase space measure is
preserved along the corresponding ray trajectories
\cite{Migliore2026Symplectic,Gradoni_SNR,Gradoni_PRA,Gradoni_AP},
such that
\begin{equation}
    \mathrm dS_{\mathrm T}\,
    \mathrm d\mathbf k_{\mathrm{t,T}}
    =
    \mathrm dS_{\mathrm R}\,
    \mathrm d\mathbf k_{\mathrm{t,R}}.
    \label{eq:phase_space_measure_conservation}
\end{equation}
This conservation holds for free-space propagation
and ideal PEC reflection, independently of the
geometrical spreading or focusing of the ray families.
Although these effects redistribute the local field
intensity, they do not alter the phase space measure
associated with the propagating rays.

For the surface-to-surface communication problem,
the complete visible transmitting-side phase space is
$\Gamma_{\mathrm{vis,T}}
=\{(\mathbf r_{\mathrm T},\mathbf k_{\mathrm{t,T}}):
\mathbf r_{\mathrm T}\in S_{\mathrm T},\,
\mathbf k_{\mathrm{t,T}}\in
\mathcal K_{\mathrm{vis},\mathrm T}(\mathbf r_{\mathrm T})\}$.
This four-dimensional phase space is sampled at
$N_{\mathrm r}$ points, each specifying an initial
position and propagation direction from which the
corresponding ray is traced through the environment.
The visibility factor
$\chi_{\mathrm{TR}}(\mathbf r_{\mathrm T,i},
\mathbf k_{\mathrm{t,T},i})$
is assigned unity if the ray reaches $S_{\mathrm R}$
through a physically admissible path, satisfying
the prescribed reflection and blockage conditions,
and zero otherwise.
The accepted initial states identify the accessible
phase space support $\Gamma_{\mathrm{TR}}$ and the
corresponding spectral region
$\mathcal K_{\mathrm{TR}}(\mathbf r_{\mathrm T})$
at each transmitting position.

Each sampled state is assigned a quadrature weight
$w_i$ representing its associated phase space measure.
When the numerical directional coordinates are not
uniform in tangential wavevector space, the
corresponding Jacobian is included in $w_i$.
The continuous phase space volume in
\eqref{eq:general_phase_space_volume} can therefore
be approximated by
\begin{equation}
    V_{\mathrm{PS}}
    \simeq
    \sum_{i=1}^{N_{\mathrm r}}
    w_i
    \chi_{\mathrm{TR}}(\mathbf r_{\mathrm T,i},
    \mathbf k_{\mathrm{t,T},i}).
    \label{eq:ray_tracing_phase_space}
\end{equation}
For sufficiently dense sampling, this quadrature
converges to the accessible phase space volume.
No additional weighting by the field amplitude
is introduced, since the geometrical mode count
depends on the accessible phase space support
rather than the coupling strength of individual
trajectories.
The traced rays are therefore quadrature samples
of phase space rather than individual communication
modes.

\begin{figure}
    \centering
    \includegraphics[width=\linewidth]{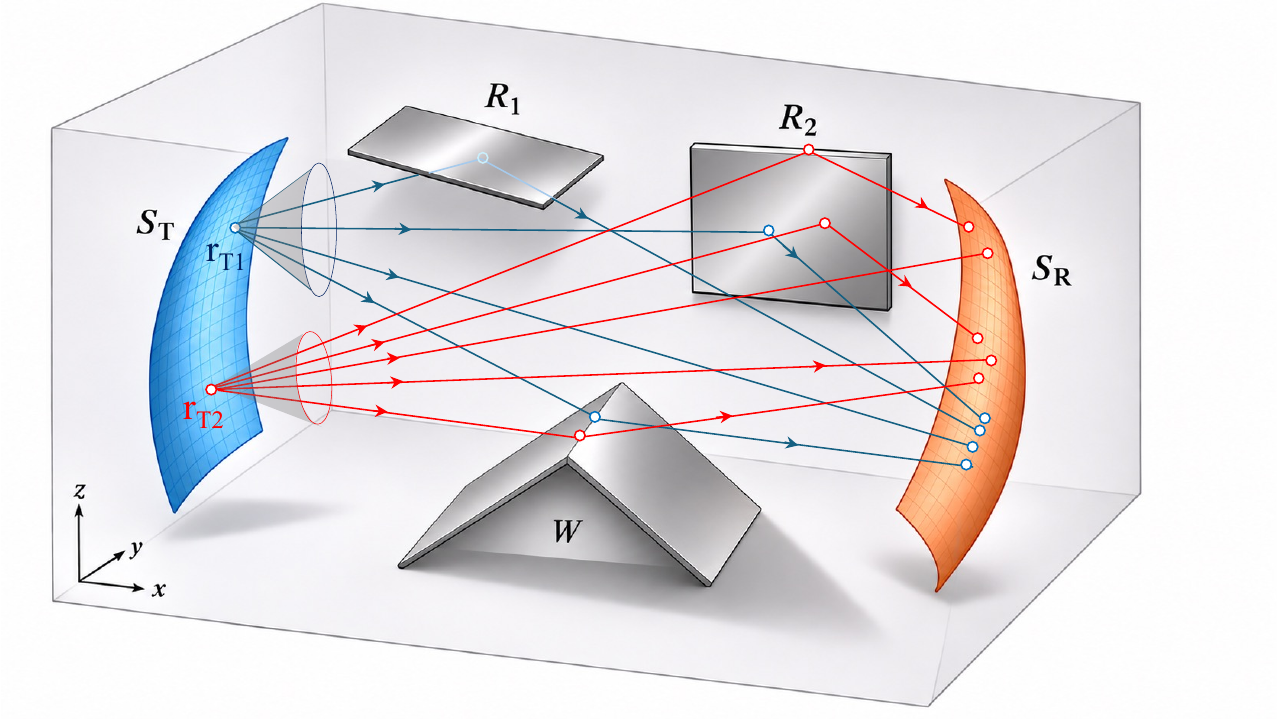}
    \caption{Schematic illustration of three-dimensional phase space ray tracing between the transmitting and receiving surfaces, \(S_{\mathrm T}\) and \(S_{\mathrm R}\), in an environment containing two planar PEC reflectors, \(R_1\) and \(R_2\), and a wedge \(W\). Edge diffraction contributions are not included in the present formulation. Two families of rays, distinguished by color, originate from different transmitting positions, \(\mathbf r_{\mathrm{T}1}\) and \(\mathbf r_{\mathrm{T}2}\), and span different initial propagation directions. The trajectories reach different points on the receiving surface through direct propagation or interactions with the reflecting environment. Each launched ray represents a sample of the transmitting-side phase space, specified by its initial position and tangential wavevector \((\mathbf r_{\mathrm T},\mathbf k_{\mathrm{t,T}})\). Rays reaching \(S_{\mathrm R}\) identify the accessible phase space support \(\Gamma_{\mathrm{TR}}\), whose volume is evaluated by integrating the corresponding phase space measure, rather than by weighting individual trajectories according to their field amplitudes.}
    \label{fig:Rays}
\end{figure}

An illustration of the mechanism is presented in Fig. \ref{fig:Rays}. It is worth noting that the present formulation is restricted to GO trajectories involving free-space propagation and specular reflections. Diffractive contributions, such as those generated by the edges of PEC wedges (see ``W'' in Fig. \ref{fig:Rays}), are therefore not included. Nevertheless, the ray tracing framework could be extended to account for such contributions, for instance through the Uniform Theory of Diffraction (UTD) \cite{UTD1, UTD2}. This would require incorporating the corresponding diffracted trajectories and their propagation directions into the characterization of the accessible phase space support.  

In reflecting environments, each ray is propagated
through the corresponding specular reflections
and contributes to the accessible phase space if
it reaches $S_{\mathrm R}$.
The incident and
reflected wavevectors at a reflecting surface are denoted by
$\mathbf k_{\mathrm i}$ and $\mathbf k_{\mathrm r}$, while
$\hat{\mathbf n}$ is the unit surface normal. Specular reflection gives
\begin{equation}
    \mathbf k_{\mathrm r}
    =
    \mathbf k_{\mathrm i}
    -
    2(\mathbf k_{\mathrm i}\cdot\hat{\mathbf n})
    \hat{\mathbf n}.
    \label{eq:specular_reflection}
\end{equation}
This relation applies to both planar and curved
PEC reflectors, using the local surface normal
at each reflection point.
The reflection preserves the wavevector magnitude
and the phase space measure while redirecting
the propagation direction
\cite{Migliore2026Symplectic,vanGestel2021Liouville}.

\subsection{Mutual Shadow Evaluation via Image Construction}
\label{subsec:image_construction}

\begin{figure}
    \centering
    \includegraphics[width=\linewidth]{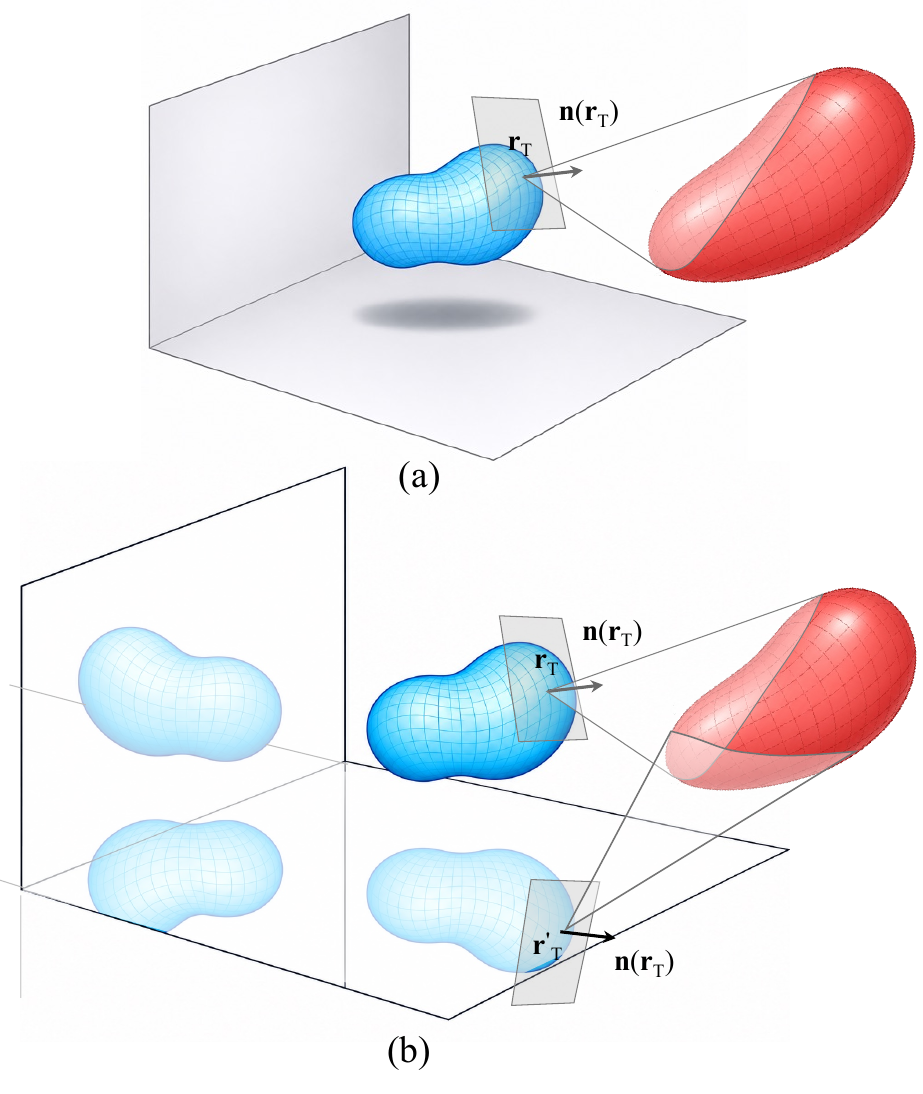}
    \caption{Geometrical interpretation of the accessible phase space and its image representation in a reflecting environment. (a) Physical configuration comprising a transmitting surface \(S_{\mathrm T}\) (blue) and a receiving surface \(S_{\mathrm R}\) (red), with a transmitting point \(\mathbf r_{\mathrm T}\) and its associated surface normal \(\mathbf n(\mathbf r_{\mathrm T})\). The connecting lines illustrate the geometrical visibility between the transmitting and receiving surfaces. (b) Equivalent image construction for two mutually perpendicular planar PEC reflectors. The physical transmitting surface (solid blue) generates three virtual images (transparent blue) through successive specular reflections. The corresponding image points and surface orientations are obtained by reflecting the physical geometry across the reflector planes. Each image defines a potential family of unfolded propagation trajectories, which contributes to the accessible phase space support only when the corresponding folded paths are physically admissible. The accessible support is therefore determined by the union of the visible contributions associated with the physical surface and its reflected images.}
    \label{fig:Images}
\end{figure}

For planar PEC reflectors, the reflected trajectories
can equivalently be described through the image
construction. 
Under the equivalent image representation, the
image electric currents $\mathbf J_{\mathrm{im}}$
and magnetic currents $\mathbf M_{\mathrm{im}}$
corresponding to the currents supported on the
transmitting aperture can be constructed. The geometrical interpretation of the image construction is illustrated in Fig. \ref{fig:Images}, relevant to a case in which the transmitting surface is within a corner. Starting from the physical transmitting and receiving surfaces in Fig. \ref{fig:Images}a, the introduction of two mutually perpendicular planar PEC reflectors leads to three virtual images of the transmitting surface, as shown in Fig. \ref{fig:Images}b. These images provide an equivalent representation of the reflected ray families, whose physically admissible contributions determine the accessible phase space support.
Defining the spatial reflection dyadic as
$\mathbf R_{\hat{\mathbf n}}
=\mathbf I-2\hat{\mathbf n}\hat{\mathbf n}$,
the image currents satisfy
$\mathbf J_{\mathrm{im}}
=-\mathbf R_{\hat{\mathbf n}}\cdot\mathbf J$
and
$\mathbf M_{\mathrm{im}}
=\mathbf R_{\hat{\mathbf n}}\cdot\mathbf M$.
Although the electric and magnetic currents undergo
different vectorial transformations, they correspond
to the same image geometry and reflected ray
trajectories.
These transformations therefore do not alter
the underlying geometrical phase space support
or mutual shadow area when both electric and
magnetic equivalent currents are retained. The accessible support is obtained from the union of the contributions associated with the physical aperture and its reflected images, as
$\Gamma_{\mathrm{TR}}
=\bigcup_{\alpha}\Gamma_{\mathrm{TR},\alpha}$.
Here, $\alpha$ indexes the distinct physically
admissible propagation sequences and their associated
ray families.
The direct family is identified by $\alpha=0$,
when included, while $\alpha>0$ identifies the
reflected families.
Each reflected image contributes only where the
corresponding unfolded trajectory maps back to
a physically realizable path connecting
$S_{\mathrm T}$ to $S_{\mathrm R}$ through the
prescribed reflector sequence.
For finite reflectors, every specular reflection
point must lie on the corresponding physical surface
and the prescribed reflection sequence must be
completed.
In non-convex configurations, trajectories are
excluded if they re-intersect the reflecting
structure, are redirected back toward the transmitting
region, or are otherwise blocked.
These conditions are incorporated into the visibility
factor $\chi_{\mathrm{TR}}$, which selects the
trajectories belonging to the accessible support.

The image construction thus identifies the same physically realizable ray families as those sampled by the ray tracing procedure in \eqref{eq:ray_tracing_phase_space}. In the unfolded geometry, the mutual shadow area can also be evaluated between the receiving aperture and the physical transmitting aperture or its reflected images, as $A_{\mathrm{TR}}=\sum_{\alpha}A_{\mathrm{TR},\alpha}$, provided that the admissible ray families occupy non-overlapping phase space regions and satisfy
the visibility constraints described above. The corresponding leading-order spatial NDoF then follows from \eqref{eq:DoF_shadow}.

The paraxial approximation developed in
Section~\ref{subsec:paraxial} is also useful for
evaluating higher-order image contributions.
For reflected-image families whose image--receiver
separation increases with reflection order,
the angular spread decreases, and the corresponding
mutual shadow contributions increasingly approach
the paraxial estimate in
\eqref{eq:paraxial_mutual_shadow},
subject to the prescribed visibility restrictions. This behavior is illustrated in the parallel-plate waveguide example of Section~\ref{subsec:ppwg},
where the paraxial approximation also establishes the asymptotic decay of higher-order image contributions.

\section{Numerical Validation in Reflecting Environments}
\label{sec:numerical_validation}

The following numerical examples specialize the general surface formulation introduced in Section~\ref{subsec:channel_operator} to planar transmitting and receiving apertures, as commonly
considered in communication systems. A planar aperture can be represented by the enclosing surface of an infinitesimally thin prism. In the zero-thickness limit, the lateral surface vanishes,
leaving two geometrically coincident faces with opposite
outward normals. For a transmitting aperture of geometrical area
$A_{\mathrm T}$, the limiting equivalent surface therefore has area $2A_{\mathrm T}$. The two oriented faces account for radiation into opposite half-spaces, consistent with the asymptotic radiating NDoF of planar apertures supporting both
electric and magnetic surface currents ~\cite{Gustafsson2025Radiating}.

The geometrical NDoF estimates are validated against the electromagnetic channel eigenvalue spectra in four
PEC reflecting environments. The first example involves planar PEC reflectors, where propagation can be represented by a finite set of direct and reflected ray families. A parabolic reflector is then examined to investigate curved specular reflection, geometrical focusing, directional propagation selection, and feed blockage. The parallel-plate waveguide (PPWG) provides an
example of multipath propagation involving an infinite
sequence of reflected ray families. Finally, an L-shaped PPWG corridor represents a non-line-of-sight (NLoS) indoor communication scenario ~\cite{Freni_Lcorridor}, where geometrical blockage eliminates the direct LoS and communication relies on physically admissible reflected paths.

For each configuration, the accessible phase space volume is evaluated geometrically through mutual shadow integration using \eqref{eq:TR_shadow}, where applicable, or through forward phase space ray tracing using \eqref{eq:ray_tracing_phase_space}. In the latter approach, initial transmitting positions and propagation directions are sampled using scrambled
Sobol sequences, a randomized quasi-Monte Carlo method~\cite{Owen1998Scrambling}. Only trajectories satisfying the prescribed reflection and visibility conditions and reaching the receiving aperture are retained. Where both methods are applicable, they provide independent geometrical NDoF estimates through \eqref{eq:phase_space_volume_DoF} and \eqref{eq:DoF_shadow}, based on the accessible phase space volume and mutual shadow area, respectively. 

For electromagnetic validation, the channel is discretized by sampling the transmitting and receiving apertures. The discretized channel is analyzed using a matrix-free randomized SVD, with the channel responses evaluated using the appropriate Green's functions
and reflection models, coherently combining contributions
from different propagation paths. The squared dominant singular values give the corresponding eigenvalues of the discretized
channel correlation matrix. The probing subspace has a dimension of at least $3N_{\mathrm A}$, following the procedure
in~\cite{Gustafsson2025Shadow}.

The comparison focuses on the spectral corner, defined here as the transition between significant and rapidly decaying eigenvalues in the normalized channel eigenvalue spectrum.
The mode index of this transition is compared
with the geometrical NDoF estimate. Although ideal PEC reflections preserve the phase space measure, reflection and geometrical visibility determine which portion of the transmitted phase space support reaches the receiving aperture.
The following examples illustrate how these effects
determine the accessible phase space volume and the
resulting spatial NDoF.

\subsection{Planar PEC Reflectors}
\label{subsec:planar_pec}

Four configurations are considered using the same transmitting
and receiving apertures: free space, a single PEC plane, a
right-angle PEC corner, and a PEC trihedral. These configurations
provide a geometrical progression in which additional specularly
reflected propagation paths are introduced without changing
the physical apertures. The transmitting and receiving apertures,
indicated by $S_{\mathrm T}$ and $S_{\mathrm R}$, are parallel
rectangles lying in planes normal to $\hat{\mathbf x}$,
with dimensions $20\lambda\times15\lambda$ and
$18\lambda\times16\lambda$, respectively.
Their centers are located at Cartesian coordinates
$(8,16,14)\lambda$ and $(28,16,15)\lambda$, respectively,
giving an aperture-plane separation of $20\lambda$.

\begin{figure}
    \centering
    \includegraphics[width=1\linewidth]{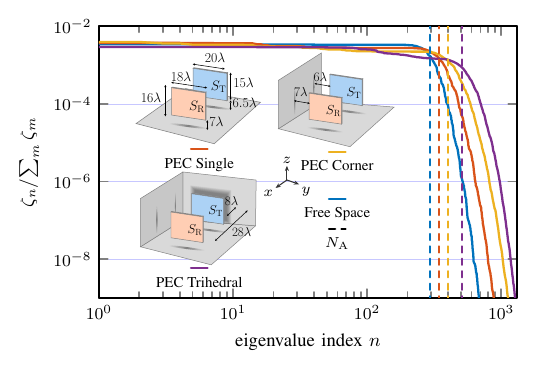}
\caption{Normalized eigenvalue spectra for the same transmitting and receiving apertures in free space, with a single PEC plane, in a right-angle PEC corner, and in a PEC trihedral. The vertical dashed lines indicate the mutual shadow NDoF predictions, with the visibility-restricted estimate used for the trihedral. The insets illustrate the physical apertures and PEC reflector configurations, with dimensions expressed in wavelengths.}
    \label{fig:planar_pec_validation}
\end{figure}

The corresponding reflector configurations are illustrated
in the insets of Fig.~\ref{fig:planar_pec_validation}. Starting from free space, the introduction of the PEC plane $z=0$ gives the single-reflector configuration. The additional plane $y=0$ forms a right-angle corner, while the plane $x=0$ completes the PEC trihedral. According to the image theorem, the number of contributions from the physical transmitting aperture and its reflected images increases from one to two, four, and eight, respectively.

The mutual shadow contributions can be evaluated numerically using \eqref{eq:TR_shadow}. For the parallel rectangular geometry considered here, a closed-form expression is also available in \cite{ehlert1993view}, with further details provided in Section~S2
of the Supplementary Material.
Accounting for the two electromagnetic polarizations,
the resulting NDoF predictions are approximately
$N_{\mathrm A}=296$, $345$, and $403$ for free space,
the single PEC plane, and the corner, respectively.

For the trihedral case, the complete eight-image construction gives $A_{\mathrm{TR}}\approx317\lambda^2$, corresponding to $N_{\mathrm A}\approx634$. An alternative geometrical interpretation arises when the physical transmitting
aperture is treated as an opaque obstacle. In this case,
the portions of the four $x$-reflected ray families
intercepted by the transmitting aperture are excluded
from the mutual shadow calculation. The resulting
visibility-restricted area is approximately $255\lambda^2$,
giving $N_{\mathrm A}\approx510$. These two estimates
represent different assumptions concerning transmitter
blockage and are compared against the same electromagnetic
channel.

The electromagnetic channels are constructed by applying the image theorem. The equivalent electric and magnetic currents on each image are determined from those on the physical transmitting aperture according to the PEC image
relations described in Section~\ref{subsec:image_construction}.
The fields radiated by the physical aperture and its images are evaluated using the full free-space dyadic Green's functions and combined coherently to form the electromagnetic channel. The image currents are determined by the physical aperture currents and are not independent excitations. For the trihedral, all eight image contributions are retained in the channel matrix without imposing an explicit transmitter-blockage condition.

Fig.~\ref{fig:planar_pec_validation} compares the normalized eigenvalue spectra with the mutual shadow NDoF predictions. The mutual shadow area estimations provide good predictions of the spectral corners \cite{Gustafsson2025Shadow}, beyond which the eigenvalues decay rapidly.
The progressive introduction of PEC reflectors shifts these corners toward higher mode indices, consistently with the increase in the mutual shadow area. For the trihedral, the visibility-restricted estimate
$N_{\mathrm A}\approx510$, indicated by the violet vertical dashed line, provides a better prediction of the spectral corner than the complete-image estimate $N_{\mathrm A}\approx634$. This result highlights the importance of geometrical visibility in accurately estimating the accessible
phase space support and the resulting spatial NDoF.

\subsection{Parabolic Reflector}
\label{subsec:parabolic_reflector}

\begin{figure}
    \centering
    \includegraphics[width=1\linewidth]{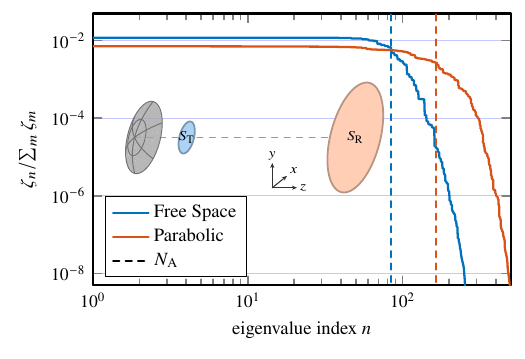}
    \caption{Normalized eigenvalue spectra of the direct
    free-space channel and the reflector-mediated parabolic
    PEC channel. The latter is evaluated using the PO
    approximation, retaining only reflected propagation
    and accounting for transmitter blockage.
    The vertical dashed lines indicate the corresponding
    mutual shadow NDoF predictions. The inset illustrates
    the transmitting aperture $S_\mathrm T$, the parabolic reflector,
    and the receiving aperture $S_\mathrm R$.}
    \label{fig:parabolic_spectra}
\end{figure}

A parabolic PEC reflector is then considered to examine the phase space formulation in the presence of curved specular reflection and geometrical focusing. The configuration, as illustrated in the inset of Fig.~\ref{fig:parabolic_spectra}, consists of a circular transmitting aperture serving as the feed and centered at the reflector focus, a PEC
paraboloid with focal length $F=14.7\lambda$ and diameter $25.2\lambda$, and a circular receiving aperture located along the reflector axis. The transmitting and receiving aperture diameters are $11.2\lambda$ and $38\lambda$, respectively. Taking the focus as the coordinate origin,
the receiving aperture is centered at $z=47.6\lambda$.

Following the conventional reflector-antenna model,
a one-sided propagation convention is adopted for
the feed. Only radiation directed toward the reflector
is retained, while radiation into the opposite half-space,
including the direct transmitter-to-receiver contribution,
is disregarded. Both equivalent electric and magnetic
surface currents remain independent in the electromagnetic
channel formulation. Furthermore, following the commonly adopted first-order
reflection approximation, the transmitting aperture is
treated as an opaque obstacle to the returning reflected
field. Reflected trajectories intercepted by the aperture
are excluded, assuming absorption and neglecting
secondary scattering and edge diffraction.

Unlike planar PEC reflectors, the parabolic reflector has a continuously varying surface normal. Following \eqref{eq:TR_shadow}, the mutual shadow area associated with the reflected propagation is
evaluated as
\begin{equation}
    A_{\mathrm{TR}}
    =
    \int_{S_{\mathrm T}}\!\int_{S_{\mathrm P}}
    \frac{
    [\hat{\mathbf n}_{\mathrm T}\cdot\widehat{\mathbf R}]_+
    [-\hat{\mathbf n}_{\mathrm P}\cdot\widehat{\mathbf R}]_+
    }{R^2}
    \chi_{\mathrm R}\chi_{\mathrm{clear}}\,
    \mathrm dS_{\mathrm P}\,\mathrm dS_{\mathrm T},
    \label{eq:parabolic_mutual_shadow}
\end{equation}
where $S_{\mathrm P}$ denotes the curved reflector surface,
$R=|\mathbf r_{\mathrm P}-\mathbf r_{\mathrm T}|$, and $\widehat{\mathbf R}=(\mathbf r_{\mathrm P}
-\mathbf r_{\mathrm T})/R$ is the incident direction
connecting a transmitting point $\mathbf r_{\mathrm T}$
to a reflector point $\mathbf r_{\mathrm P}$.
Here, $\hat{\mathbf n}_{\mathrm T}$ is the unit
normal to the transmitting aperture oriented
toward the reflector, while
$\hat{\mathbf n}_{\mathrm P}(\mathbf r_{\mathrm P})$
is the local reflector normal oriented toward
the feed. The latter varies along
the curved surface and determines the reflected
direction according to \eqref{eq:specular_reflection}.
The visibility factors $\chi_{\mathrm R}$ and
$\chi_{\mathrm{clear}}$ equal unity when the corresponding
reflected propagation path intersects the finite receiving
aperture and avoids the transmitting aperture, respectively,
and vanish otherwise.

Eq.~\eqref{eq:parabolic_mutual_shadow} thus evaluates the mutual shadow area associated with the accessible reflected phase space by integrating over pairs of points on the transmitting aperture and the reflector. The same support can alternatively be evaluated by directly sampling positions and propagation
directions in the transmitting phase space, using the
phase space ray tracing method introduced in \eqref{eq:ray_tracing_phase_space}. In this approach,
rays are sampled directly in the transmitting phase
space, propagated toward the paraboloid, and reflected
according to the local surface normals. The same
receiver-interception and transmitter-blockage conditions
are then applied to determine the accessible phase space
volume. For the present geometry, both the mutual shadow area integration in ~\eqref{eq:parabolic_mutual_shadow} and phase space ray tracing predict $N_{\mathrm A}\approx163$, with a relative difference below $0.01\%$ before rounding, confirming the consistency of the two geometrical approaches.

The electromagnetic channel is constructed using the
physical-optics (PO) approximation \cite{balanis2012antenna}. The incident magnetic field $\mathbf H_{\mathrm{inc}}$ on the reflector is
first evaluated from the independent electric and magnetic
currents on $S_{\mathrm T}$ using the full free-space
dyadic Green's functions. Since the tangential electric
field vanishes on the PEC surface, the corresponding
equivalent magnetic current is zero, and the reflector
field can be represented by an equivalent electric
surface current determined by the tangential magnetic
field. Under the PO approximation, the local planar
reflection gives
\begin{equation}
    \mathbf J_{\mathrm{PO}}
    =2\hat{\mathbf n}_{\mathrm P}\times
    \mathbf H_{\mathrm{inc}},
    \label{eq:parabolic_po_current}
\end{equation}
where $\hat{\mathbf n}_{\mathrm P}$ is the local reflector
normal directed toward the incident field. The factor
of two follows from the local PEC image construction,
where the incident and reflected tangential magnetic
fields add at the reflector surface. The field radiated by $\mathbf J_{\mathrm{PO}}$ is subsequently evaluated at $S_{\mathrm R}$, excluding
propagation segments intercepted by the transmitting
aperture, consistent with the previous formulation. 

Fig.~\ref{fig:parabolic_spectra} compares the
normalized eigenvalue spectrum of the parabolic-reflector
channel with that of the direct free-space channel,
included as a separate reference. The mutual shadow
prediction $N_{\mathrm A}\approx163$ provides a good
estimate of the spectral corner of the reflector-mediated
channel, beyond which the eigenvalues decay rapidly.

Compared with the direct free-space reference, for which
$N_{\mathrm A}\approx84$, the parabolic reflector provides
a substantially larger accessible phase space support.
Its curved surface redirects radiation from the focal
region into an approximately collimated outgoing field,
improving its accessibility to the receiving aperture. To further illustrate the role of reflector curvature, two additional geometrical benchmarks are considered. Replacing the paraboloid with a finite planar PEC reflector of the same projected diameter $25.2\lambda$, positioned
at the paraboloid vertex and subject to the same feed-blockage
condition, gives $N_{\mathrm A}\approx16$ for the
reflected propagation, compared with $N_{\mathrm A}\approx163$ for the parabolic reflector. As another reference, replacing the transmitting aperture with a larger circular aperture of diameter $25.2\lambda$,
matching the projected reflector aperture and centered
at the original focal point, gives a free-space estimate
of $N_{\mathrm A}\approx409$ for the same receiver.
These comparisons highlight that the parabolic reflector
does not introduce additional independent source DoF,
but redirects the existing phase space support of the transmitter to increase its accessibility to the receiver, while preserving
the underlying phase space measure.

\subsection{Parallel-Plate Waveguide}
\label{subsec:ppwg}

\begin{figure}
    \centering
    \includegraphics[width=1\linewidth]{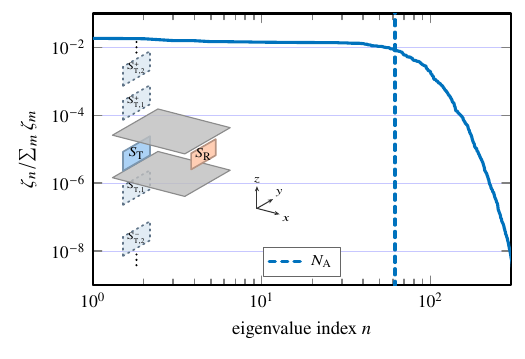}
    \caption{Normalized electromagnetic channel eigenvalue
    spectrum for the PPWG, with image contributions
    retained through reflection order $P=8$.
    The vertical dashed line indicates the accumulated
    mutual shadow NDoF prediction $N_{\mathrm A}$.
    The inset illustrates the transmitting and receiving
    apertures, the parallel PEC plates, and the
    unfolded images $S^{\pm}_{\mathrm{T},p}$ for the first two reflection orders.}
    \label{fig:ppwg_spectra}
\end{figure}

The parallel-plate waveguide (PPWG) is subsequently considered to investigate the mutual shadow formulation in an environment supporting an infinite number of reflected propagation
paths. The configuration consists of two infinite parallel PEC plates separated by $h=24\lambda$, with transmitting and receiving apertures lying in parallel planes normal
to $\hat{\mathbf x}$, as illustrated in the inset of Fig.~\ref{fig:ppwg_spectra}. The transmitting and receiving apertures have dimensions $10\lambda\times7.5\lambda$ and
$9\lambda\times8\lambda$, with centers at $(0,0,-2)\lambda$ and $(10,5,0)\lambda$, respectively, giving an axial separation of $d=10\lambda$. The plate separation is chosen sufficiently large to accommodate both apertures and avoid imposing a restrictive modal cutoff on the propagating spatial modes relevant to the communication channel between the transmitting and receiving apertures \cite{Gus24_DoF_CM}.

Successive reflections between the plates generate two
infinite families of transmitting images. Taking the
plate normal along $\hat{\mathbf z}$ and the waveguide
midplane at $z=0$, the plates are located at
$z=\pm h/2$. The image positions of a source point
with coordinate $z'$ at reflection order $p\geq1$ are
\begin{equation}
    z'_{p,\pm}=(-1)^p z'\pm ph.
    \label{eq:ppwg_image_position}
\end{equation}
Each positive reflection order introduces two
images, $S^+_{\mathrm{T},p}$ and $S^-_{\mathrm{T},p}$, in addition to the
physical transmitting aperture $S_\mathrm T$ at $p=0$. Their equivalent electric and magnetic currents follow the PEC image transformations introduced in
Section~\ref{subsec:image_construction}. 

Since the transmitting images are parallel to the receiving aperture, their mutual shadow contributions can be evaluated numerically using \eqref{eq:TR_shadow}
or analytically using the closed-form expression in \cite{ehlert1993view}.
The cumulative mutual shadow area, considering
reflections up to a maximum order $P$, is
\begin{equation}
    A_{\mathrm{TR}}^{(P)}
    =
    A_0+\sum_{p=1}^{P}(A_{p,+}+A_{p,-}),
    \label{eq:ppwg_cumulative_shadow}
\end{equation}
where $A_0$ is the direct contribution and
$A_{p,\pm}$ correspond to the two image families.
The associated two-polarization NDoF estimate is
$N_{\mathrm A}^{(P)}=2A_{\mathrm{TR}}^{(P)}/\lambda^2$.

\begin{figure}
    \centering
    \includegraphics[width=1\linewidth]{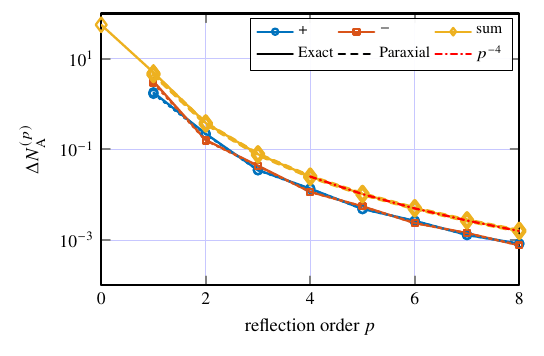}
    \caption{Incremental two-polarization NDoF contributions
    $\Delta N_{\mathrm A}^{(p)}$ versus reflection order
    $p$ in the PPWG. The two image families ($+$ and $-$)
    and their sum are compared using exact mutual shadow
    integration (solid lines) and the paraxial
    approximation (dashed lines). The dash-dotted line
    indicates the asymptotic $p^{-4}$ decay.}
    \label{fig:ppwg_shadow}
\end{figure}

An important property of this configuration is the
asymptotic decay of the higher-order mutual shadow
contributions. Applying the paraxial approximation
of \eqref{eq:paraxial_parallel_shadow} to each
reflected image gives
\begin{equation}
    A_{p,\pm}^{\mathrm{par}}
    =
    \frac{A_{\mathrm T}A_{\mathrm R}d^2}{R_{p,\pm}^4}
    \approx
    \frac{A_{\mathrm T}A_{\mathrm R}d^2}{p^4h^4},
    \label{eq:ppwg_paraxial_image}
\end{equation}
where $R_{p,\pm}$ is the center-to-center distance
between the corresponding transmitting image and the
receiving aperture, and the final approximation applies
at sufficiently large reflection orders.
For these parallel apertures, each projection factor
in the mutual shadow integral of \eqref{eq:TR_shadow} scales as
$d/R_{p,\pm}$, yielding the $p^{-4}$ decay with increasing reflection order. Higher-order images correspond to longer and increasingly oblique
propagation paths, contributing progressively smaller mutual shadow areas despite lossless propagation and reflection. Consequently, the accumulated
mutual shadow area converges despite the infinite number of reflected images.

Fig.~\ref{fig:ppwg_shadow} compares the 
incremental mutual shadow contributions with the
paraxial estimates for both image families.
The agreement improves with increasing reflection
order, and their combined contribution approaches
the predicted $p^{-4}$ asymptotic behavior.
The image-based calculation is also independently
verified using the forward phase space ray tracing
procedure of \eqref{eq:ray_tracing_phase_space}.

The electromagnetic channel is constructed by
coherently combining the fields radiated by the
physical transmitting aperture and its PEC images.
The image expansion is truncated at $P=8$, as
Fig.~\ref{fig:ppwg_shadow} shows that the combined
mutual shadow contribution at this order has decreased
to $\Delta N_{\mathrm A}^{(8)}$ of the order of $10^{-3}$,
with higher-order contributions decreasing further.
Extrapolating the observed $p^{-4}$ decay, the total
contribution from all neglected higher-order images
is estimated to be below $4\times10^{-3}$, negligible relative to the cumulative NDoF of the electrically large configuration considered.

Fig.~\ref{fig:ppwg_spectra} compares the normalized
eigenvalue spectrum of the resulting electromagnetic
channel with the cumulative mutual shadow prediction
$N_{\mathrm A}\approx62$, which accurately estimates
the spectral corner beyond which the eigenvalues
decay rapidly. This example demonstrates that
an unbounded number of geometrically distinct
reflection paths does not imply an unbounded number
of accessible communication modes. Instead, the
spatial NDoF remains governed by their finite
cumulative phase space measure.

\subsection{L-Shaped Parallel-Plate Corridor}
\label{subsec:l_corridor}

\begin{figure}
    \centering
    \includegraphics[width=1\linewidth]{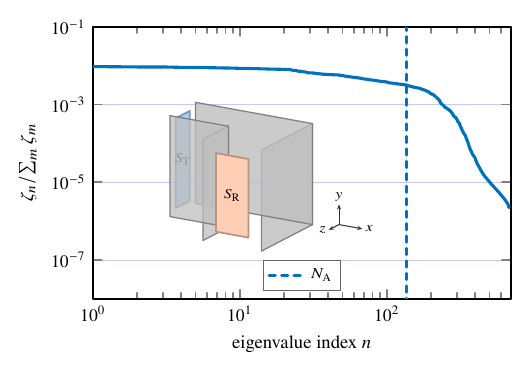}
    \caption{Normalized electromagnetic channel
    eigenvalue spectrum for the L-shaped PEC
    corridor, with reflected contributions
    retained through order $P=24$.
    The vertical dashed line indicates the
    phase space NDoF prediction $N_{\mathrm A}$.
    The inset illustrates the NLoS corridor
    geometry and the orthogonally oriented
    transmitting and receiving apertures.}
    \label{fig:l_corridor_spectra}
\end{figure}

Finally, an L-shaped PEC parallel-plate corridor is considered to
investigate the phase space formulation in an NLoS
reflecting environment, where geometrical visibility
plays a central role in determining the accessible
communication modes. The corridor has a width
$h=18\lambda$ and extends $36\lambda$ along both
the $x$ and $z$ directions, with PEC walls extending
infinitely along $\hat{\mathbf y}$.
The transmitting and receiving apertures are located
at the two ends of the corridor, in mutually orthogonal
planes normal to $\hat{\mathbf x}$ and
$\hat{\mathbf z}$, respectively, as illustrated in
Fig.~\ref{fig:l_corridor_spectra}.
The transmitting aperture has dimensions
$32\lambda\times10\lambda$, while the receiving
aperture has dimensions $28\lambda\times10\lambda$.
Their centers are located at
$(0,0,9)\lambda$ and $(27,0,36)\lambda$,
respectively. Owing to the NLoS geometry,
there is no direct LoS between the
two apertures, and communication relies entirely
on reflected propagation around the corridor bend.

The NLoS geometry of the L-shaped corridor imposes complicated visibility restrictions on the propagation paths associated with transmitting images, making direct image-based mutual shadow integration less convenient. The accessible phase space support is therefore estimated using the forward ray tracing formulation of \eqref{eq:ray_tracing_phase_space}, retaining only trajectories that reach the finite receiving aperture. 

\begin{figure}
    \centering
    \includegraphics[width=1\linewidth]{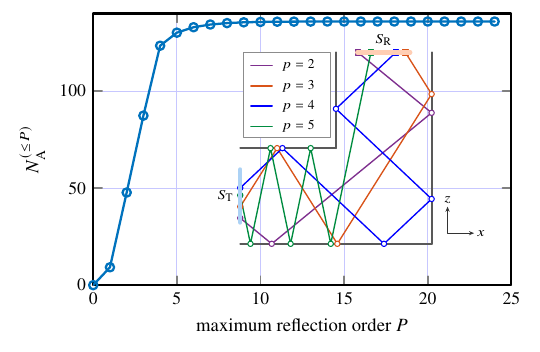}
    \caption{Accumulated spatial NDoF
    $N_{\mathrm A}^{(\leq P)}$ in the L-shaped
    PEC corridor as a function of the maximum
    reflection order $P$. The inset shows
    representative physically admissible ray
    trajectories connecting the transmitting
    and receiving apertures, with reflection
    orders $p=2,3,4,5$.}
    \label{fig:l_corridor_shadow}
\end{figure}

Fig.~\ref{fig:l_corridor_shadow} shows the
accumulated estimate $N_{\mathrm A}^{(\leq P)}$
as the maximum reflection order $P$ increases
to 24, together with representative trajectories
connecting the two apertures. The lowest admissible
reflection orders provide the dominant contributions,
while higher-order families contribute progressively
smaller portions of the accessible phase space.
The accumulated estimate converges toward
$N_{\mathrm A}\approx136$. This demonstrates that
the geometrical visibility restrictions of the
corridor yield a finite accessible phase space
measure despite the increasing number of possible
reflection sequences.

For the electromagnetic validation, the channel
is constructed using the full free-space dyadic
Green's functions, with successive PEC reflections
represented through transmitting images. However, an unfolded connection between an image and the receiver does not necessarily correspond to a physically admissible trajectory. For each candidate image, the corresponding ray
is traced in the physical corridor and retained
only if it follows the prescribed sequence of
reflections on the finite PEC walls and reaches
the intended receiving point. The appropriate PEC current transformations are applied to the validated images, whose contributions
are combined coherently in the channel matrix.
The channel is evaluated through reflection order
$P=24$ using the randomized-SVD procedure.

Fig.~\ref{fig:l_corridor_spectra} compares the
normalized eigenvalue spectrum of the resulting
electromagnetic channel with the phase space
prediction $N_{\mathrm A}\approx136$, which
accurately identifies the spectral corner even
in this NLoS configuration, where geometrical
visibility excludes inaccessible states from
$\Gamma_{\mathrm{TR}}$ rather than merely
reducing their field amplitudes. The spatial
NDoF is thus governed by the phase space measure
of physically admissible propagation paths.

\subsection{Discussion}
\label{subsec:numerical_discussion}

The preceding examples demonstrate that additional
reflection paths do not necessarily introduce
independent spatial communication modes.
The equivalent currents of each reflected image
are determined by the physical transmitting currents
through the corresponding PEC transformations,
and their fields are combined coherently into a
single channel operator. While reflections can
enlarge the accessible phase space support,
the resulting NDoF is governed by its total measure
rather than the number of distinct reflection paths.

Free-space propagation and ideal specular reflection
preserve the canonical phase space measure, as
established in Section~\ref{subsec:GO_interpretation}.
The mutual shadow area quantifies the portion of
this conserved phase space measure that connects
the transmitting and receiving apertures, subject
to the geometrical visibility constraints of
the propagation environment. The planar PEC configurations illustrate
the additional reflected paths introduced by
successive reflections, while the parabolic reflector
redistributes the accessible propagation directions
through focusing and is subject to feed blockage.
In the PPWG, increasingly higher reflection orders
contribute progressively smaller phase space areas,
whereas the L-shaped corridor demonstrates the
importance of geometrical visibility in the absence
of a direct LoS. In all cases, the accessible phase space support
is determined by the geometrically admissible ray trajectories connecting the transmitting and receiving apertures.

The sidedness of the equivalent aperture, introduced at the beginning of this section, should be distinguished
from geometrical blockage. For the parabolic reflector, the one-sided propagation convention restricts the
initial phase space to reflector-facing directions, whereas feed blockage excludes reflected trajectories
intercepted by the transmitting aperture. Both restrictions reduce the accessible phase space support without affecting the two electromagnetic polarizations counted for each retained propagation state.

Finally, the agreement between the phase space
predictions and the electromagnetic eigenvalue
spectra across the considered configurations
supports the leading-order NDoF interpretation
for geometrically accessible, lossless canonical
propagation. However, the individual modal coupling strengths are not predicted by this process. Beyond ideal PEC reflections, refraction preserves the canonical phase space measure \cite{vanGestel2021Liouville}, while partially transmitting interfaces introduce angle- and polarization-dependent Fresnel coefficients, which have been incorporated into Liouville-based phase space energy transport \cite{JinWen2006,vanGestel2023Thesis}.
Such amplitude weighting can modify the coupling strengths of individual electromagnetic modes. Its implications for NDoF estimation are beyond the scope of the present PEC analysis. 

\section{Conclusion}
\label{sec:conclusion}
This work has developed a geometrical formulation for estimating the leading-order spatial NDoF of electromagnetic communication channels between electrically large transmitting and receiving surfaces in reflecting environments. The full-vector electromagnetic channel is formulated using equivalent electric and magnetic surface currents, while its leading-order NDoF is estimated from the accessible phase space volume. Through two geometrical projection Jacobians, this volume is related to the mutual shadow area, yielding the two-polarization NDoF estimate $N_{\mathrm{DoF}}\simeq2A_{\mathrm{TR}}/\lambda^2$. Extending the free-space mutual shadow formulation to reflecting environments, the proposed approach retains its applicability to nonparaxial propagation while accounting for specular reflections and geometrical
visibility constraints through mutual shadow integration and phase space ray tracing.

The geometrical NDoF estimates have been validated against the electromagnetic channel eigenvalue spectra in four PEC reflecting environments, including planar reflectors, a parabolic reflector, a parallel-plate waveguide, and an NLoS L-shaped parallel-plate corridor. The predictions consistently identify the spectral corners, including in configurations involving focusing, multiple reflections, and geometrical blockage. These results demonstrate that PEC reflections can redirect propagation and increase the phase space volume accessible to the receiver, while the resulting NDoF is governed by the total accessible volume rather than the number of distinct propagation paths.

The proposed approach provides a direct and physically interpretable geometrical estimate of the leading-order spatial NDoF through mutual shadow integration or phase space ray tracing, without requiring the construction and decomposition of the full electromagnetic channel. The agreement across the considered configurations
supports its applicability to communication between electrically large apertures in PEC reflecting environments.

\appendices
\section{Free-Space Green Functions}
\label{app:dyadic_green}

The numerical electromagnetic channel is constructed using the full impedance-normalized electric- and
magnetic-current dyadic Green's functions, consistent with \eqref{eq:channel_operator}.
For a source point $\mathbf r'$ and an observation point $\mathbf r$, define
$\mathbf R=\mathbf r-\mathbf r'$, $R=|\mathbf R|$, and $\hat{\mathbf R}=\mathbf R/R$.
Using the $\mathrm e^{\mathrm j\omega t}$ phasor convention, with time dependence suppressed,
the normalized electric-current dyadic Green's function \cite{tai1994dyadic} is 
\begin{equation}
\mathbf G^{\mathrm{EJ}}(\mathbf R)
=
-\mathrm j k_0
\left(
\mathbf I+\frac{\nabla\nabla}{k_0^2}
\right)
\frac{\mathrm e^{-\mathrm j k_0R}}{4\pi R}.
\label{eq:full_dyadic_green_EJ}
\end{equation}
Unlike the conventional electric-current Green's function, the factor $\eta_0$ is absorbed into
the normalized source $\eta_0\mathbf J$ in \eqref{eq:channel_operator}, leaving $\mathbf G^{\mathrm{EJ}}$ with units of inverse area.
The corresponding magnetic-current dyadic
Green's function is
\begin{equation}
\mathbf G^{\mathrm{EM}}(\mathbf R)
=
-\nabla\times
\left(
\mathbf I
\frac{\mathrm e^{-\mathrm j k_0R}}{4\pi R}
\right).
\label{eq:full_dyadic_green_EM}
\end{equation}
The normalized magnetic-field Green's functions
follow from electromagnetic duality as
\begin{equation}
\mathbf G^{\mathrm{HJ}}(\mathbf R)
=
-\mathbf G^{\mathrm{EM}}(\mathbf R),
\label{eq:full_dyadic_green_HJ}
\end{equation}
and
\begin{equation}
\mathbf G^{\mathrm{HM}}(\mathbf R)
=
\mathbf G^{\mathrm{EJ}}(\mathbf R).
\label{eq:full_dyadic_green_HM}
\end{equation}
Here, the differential operators act with respect
to the observation coordinates.
The dyadics $\mathbf G^{\mathrm{EM}}$ and
$\mathbf G^{\mathrm{HJ}}$ contain the $1/R$
and $1/R^2$ contributions, whereas
$\mathbf G^{\mathrm{EJ}}$ and
$\mathbf G^{\mathrm{HM}}$ additionally contain
the $1/R^3$ contribution.

To construct the surface-to-surface channel,
the dyadic Green's functions are projected onto
the local tangent planes of the transmitting
and receiving surfaces.
The corresponding tangential projection dyadics are
$\mathbf P_{\mathrm T}
=\mathbf I-\hat{\mathbf n}_{\mathrm T}
\hat{\mathbf n}_{\mathrm T}$ and
$\mathbf P_{\mathrm R}
=\mathbf I-\hat{\mathbf n}_{\mathrm R}
\hat{\mathbf n}_{\mathrm R}$,
respectively.
Each tangential Green's function block in
\eqref{eq:G} is then obtained as
\begin{equation}
\mathbf G_{\mathrm t}^{\mathrm{AB}}
=
\mathbf P_{\mathrm R}
\cdot\mathbf G^{\mathrm{AB}}
\cdot\mathbf P_{\mathrm T},
\label{eq:tangential_green_projection}
\end{equation}
for the four source--field combinations $\mathrm{AB}=\mathrm{EJ}, \mathrm{EM}, \mathrm{HJ}, \mathrm{HM}$. The projection on the transmitting side restricts the source currents to the local tangent plane at $\mathbf r_\mathrm T$, while the receiving-side projection retains the tangential electric and magnetic field components at $\mathbf r_{\mathrm R}$ on $S_{\mathrm R}$.

\section*{Acknowledgment}

OpenAI's ChatGPT and Codex were used to assist with
manuscript preparation and numerical implementation. All AI-assisted content was reviewed and verified by the authors, who take full responsibility for the accuracy and conclusions of this work.

\bibliographystyle{IEEEtran}
\bibliography{reference.bib}

\end{document}

\title{Supplementary Material: Geometrical Estimation of Electromagnetic Degrees of Freedom in Reflecting Environments}
\author{Yanwen Chen, \IEEEmembership{Graduate Student Member, IEEE}, Stefano Maci, \IEEEmembership{Fellow, IEEE} and Mats Gustafsson, \IEEEmembership{Senior Member, IEEE}
}

\maketitle

This supplementary material provides additional
mathematical details supporting the accompanying paper.
Section~\ref{sec:slepian_landau} reviews the classical Slepian--Landau
concentration problem and its connection to the
electromagnetic channel eigenproblem.
Section~\ref{sec:parallel_rectangle_shadow} presents the closed-form expression for
the mutual shadow area of parallel rectangular
apertures used in the numerical validation.

\section{Slepian--Landau Concentration Problem}
\label{sec:slepian_landau}

The classical spatial--spectral concentration problem developed by Slepian, Pollak, and Landau provides the canonical framework for counting the number of independent functions that can be simultaneously concentrated within bounded spatial and spectral regions \cite{SlepianPollak_1,LandauPollak_3,Slepian_4,
Franceschetti2015Laundau}. 

Consider a canonical \emph{scalar} setting where an unobstructed planar transmitting aperture $S_{\mathrm T}$
radiates into a homogeneous half-space. The receiver is represented by a fixed tangential wavevector acceptance region $\mathcal K_{\mathrm{TR}}$, independent of the transmitting position, such that $\mathcal K_{\mathrm{TR}}(\mathbf r_{\mathrm T})
=\mathcal K_{\mathrm{TR}}$. In this setting, the scalar propagation operator in the homogeneous
half-space is translationally invariant along the aperture plane and
therefore reduces to a spatial convolution, which can be diagonalized
by the tangential Fourier transform.
If the receiver accepts all outward propagating directions,
$\mathcal K_{\mathrm{TR}}$ coincides with the full visible region
$\mathcal K_{\mathrm{vis},\mathrm T}$.

The classical Slepian--Landau problem considers the
concentration of a scalar field $f(\mathbf r_{\mathrm T})$ within a spatial region $S_{\mathrm T}$ and a fixed tangential wavevector region $\mathcal K_{\mathrm{TR}}$. The corresponding concentration operator is~\cite{SlepianPollak_1,LandauPollak_3,Slepian_4,
Franceschetti2015Laundau}
\begin{equation}
    \mathcal C
    =
    \chi_{S_{\mathrm T}}\mathcal F^{-1}
    \chi_{\mathcal K_{\mathrm{TR}}}\mathcal F
    \chi_{S_{\mathrm T}},
    \label{eq:concentration_operator}
\end{equation}
where $\mathcal F$ denotes the two-dimensional Fourier
transform, and $\chi_{\mathcal A}$ equals unity inside
$\mathcal A$ and zero otherwise.
The field $f(\mathbf r_{\mathrm T})$ is first restricted to $S_{\mathrm T}$ by $\chi_{S_{\mathrm T}}$ and transformed to the tangential wavevector domain, where $\chi_{\mathcal K_{\mathrm{TR}}}$ retains only the components within $\mathcal K_{\mathrm{TR}}$. The resulting spectrum is then transformed back to the spatial domain and restricted again to $S_{\mathrm T}$ by $\chi_{S_{\mathrm T}}$.
This spatial--spectral restriction provides a canonical
scalar analogue of the forward--adjoint structure of
$\mathcal H^\dagger\mathcal H$.
The corresponding integral representation is
\begin{equation}
    (\mathcal C f)(\mathbf r_{\mathrm T,2})
    =
    \int_{S_{\mathrm T}}
    C_{\mathcal K}(\mathbf r_{\mathrm T,2}-\mathbf r_{\mathrm T,1})
    f(\mathbf r_{\mathrm T,1})
    \,\mathrm dS_{\mathrm T,1},
    \label{eq:concentration_explicit}
\end{equation}
where
$C_{\mathcal K}(\boldsymbol{\rho})
=(2\pi)^{-2}\int_{\mathcal K_{\mathrm{TR}}}
\mathrm e^{\mathrm j\mathbf k_{\mathrm{t,T}}\cdot\boldsymbol{\rho}}
\,\mathrm d\mathbf k_{\mathrm{t,T}}$
is the inverse Fourier transform of
$\chi_{\mathcal K_{\mathrm{TR}}}$.

The concentration modes $\psi_n$ are obtained as the eigenfunctions of $\mathcal C$, satisfying $\mathcal C\psi_n=\mu_n\psi_n$. The spatial restriction to the finite aperture $S_{\mathrm T}$ necessarily gives rise to an unbounded Fourier spectrum $\tilde{\psi}_n=\mathcal F\psi_n$.
The eigenvalue $\mu_n$ therefore quantifies the fraction of
the total spectral energy concentrated within the accessible
region $\mathcal K_{\mathrm{TR}}$, relative to the entire
spectral domain $\mathbb R^2$. By Parseval's theorem, this gives
\begin{equation}
    \mu_n
    =
    \frac{\displaystyle\int_{\mathcal K_{\mathrm{TR}}}
    |\tilde{\psi}_n(\mathbf k_{\mathrm {t,T}})|^2
    \,\mathrm d\mathbf k_{\mathrm {t,T}}}
    {\displaystyle\int_{\mathbb R^2}
    |\tilde{\psi}_n(\mathbf k_{\mathrm {t,T}})|^2
    \,\mathrm d\mathbf k_{\mathrm {t,T}}} .
    \label{eq:concentration_eigenvalue}
\end{equation}
Hence, $0\leq\mu_n\leq1$. Eigenvalues close to unity correspond to modes whose spectral
energy is predominantly contained within
$\mathcal K_{\mathrm{TR}}$, whereas eigenvalues close to zero
indicate weak concentration within the accessible region
\cite{LandauPollak_3,Franceschetti2015Laundau}.

Since the same tangential wavevector region $\mathcal K_{\mathrm{TR}}$ is accessible at every
point of $S_{\mathrm T}$ in the present case, the accessible phase space support is
$\Gamma_{\mathrm{TR}} =S_{\mathrm T}\times\mathcal K_{\mathrm{TR}}$. Its phase space volume
is therefore
\begin{equation}
    V_{\mathrm{PS}}
    =
    \int_{S_{\mathrm T}}
    \int_{\mathcal K_{\mathrm{TR}}}
    \mathrm d\mathbf k_{\mathrm {t,T}}\,
    \mathrm dS_{\mathrm T}
    =
    |S_{\mathrm T}|\,|\mathcal K_{\mathrm{TR}}|.
    \label{eq:separable_phase_space_volume}
\end{equation}
Here,
$|S_{\mathrm T}|=\int_{S_{\mathrm T}}\mathrm dS_{\mathrm T}$ is the
physical transmit aperture area, whereas
$|\mathcal K_{\mathrm{TR}}|
=\int_{\mathcal K_{\mathrm{TR}}}
\mathrm d\mathbf k_{\mathrm {t,T}}$ is the area of the accessible region
in tangential wavevector space
\cite{Migliore2026Symplectic,Gradoni_PRA}.

As $V_{\mathrm{PS}}$ increases, the ordered concentration eigenvalues exhibit an increasingly sharp drop from values close to unity to values close to zero
\cite{LandauPollak_3,Slepian_4,Franceschetti2015Laundau}.
The concentration operator satisfies an exact
spectral sum rule. Setting
$\mathbf r_{\mathrm T,2}=\mathbf r_{\mathrm T,1}$
reduces the spatial convolution kernel in
\eqref{eq:concentration_explicit} to
$|\mathcal K_{\mathrm{TR}}|/(2\pi)^2$. Integration over
$S_{\mathrm T}$ then gives
\begin{equation}
    \sum_n\mu_n
    =
    \operatorname{Tr}\mathcal C
    =
    \frac{|S_{\mathrm T}|\,|\mathcal K_{\mathrm{TR}}|}
    {(2\pi)^2}
    =
    \frac{V_{\mathrm{PS}}}{(2\pi)^2}.
    \label{eq:concentration_trace}
\end{equation}
The trace identity alone does not determine the number of well-concentrated modes. However, the asymptotic clustering of concentration eigenvalues near zero and unity gives the leading-order \emph{scalar} mode count $V_{\mathrm{PS}}/(2\pi)^2$. The corresponding electromagnetic NDoF estimate
in the accompanying paper accounts for both
transverse polarizations and is therefore twice this scalar count.

The classical Slepian--Landau concentration operator provides a scalar reference for communication-channel eigenproblems. When the receiver is idealized as a fixed,
uniformly weighted spectral restriction, the corresponding scalar correlation operator reduces to the Slepian--Landau form. For well-separated, aligned rectangular apertures, a related reduction of the scalar Green's function communication eigenproblem is obtained under the paraxial approximation~\cite{Miller00}.

For the general electromagnetic channel, we write
schematically
$\mathcal H=\chi_{S_{\mathrm R}}\mathcal G
\chi_{S_{\mathrm T}}$,
where $\mathcal G$ represents the impedance-normalized
electromagnetic Green's function operator mapping
the tangential electric- and magnetic-current
components on the transmitting surface to the
tangential field components on the receiving surface.
The operators $\chi_{S_{\mathrm T}}$ and
$\chi_{S_{\mathrm R}}$ represent the corresponding
surface restrictions. The transmit-side
correlation operator then has the schematic structure
\begin{equation}
    \mathcal H^\dagger\mathcal H
    =
    \chi_{S_{\mathrm T}}\mathcal G^\dagger
    \chi_{S_{\mathrm R}}\mathcal G\chi_{S_{\mathrm T}}.
    \label{eq:general_correlation_structure}
\end{equation}
Its restriction--propagation--restriction structure
is analogous to that of the classical
Slepian--Landau concentration operator in
\eqref{eq:concentration_operator}~\cite{Gradoni_AP}.

Despite this structural similarity, the eigenvalues
of the two operators have different interpretations.
The classical concentration eigenvalues $\mu_n$
quantify normalized spectral concentration,
whereas the eigenvalues $\zeta_n$ of the general
electromagnetic correlation operator characterize
modal coupling strengths.
Although the latter satisfy an exact spectral
sum rule, their sum does not directly determine
the number of significant communication modes.
For the general surface-to-surface configuration,
the leading-order geometrical NDoF is instead
estimated from the generally non-separable
accessible phase space volume, as developed
in the accompanying paper.

\section{Mutual Shadow of Parallel Rectangular Apertures}
\label{sec:parallel_rectangle_shadow}

Consider two parallel rectangular apertures $S_1$ and $S_2$,
separated by a distance $h$ and bounded by
$y_p^-\leq y_p\leq y_p^+$ and
$z_p^-\leq z_p\leq z_p^+$, $p\in\{1,2\}$. Their mutual shadow is
\begin{equation}
    A_{12}
    =
    \int_{S_1}\int_{S_2}
    \frac{h^2}{
    \left[
    h^2+(y_1-y_2)^2+(z_1-z_2)^2
    \right]^2}
    \,\mathrm dS_2\,\mathrm dS_1 .
    \label{eq:parallel_rectangle_shadow_integral}
\end{equation}
A closed-form evaluation of this integral follows from the
parallel-rectangle result of Ehlert and Smith
\cite{ehlert1993view}. Define
$y_{1,1}=y_1^-$, $y_{1,2}=y_1^+$,
$y_{2,1}=y_2^-$, $y_{2,2}=y_2^+$,
$z_{1,1}=z_1^-$, $z_{1,2}=z_1^+$,
$z_{2,1}=z_2^-$, and $z_{2,2}=z_2^+$, with
$i,j,k,l\in\{1,2\}$ and $s_1=-1$, $s_2=1$.
Defining
$Y_{ik}=(y_{1,i}-y_{2,k})/h$ and
$Z_{jl}=(z_{1,j}-z_{2,l})/h$, the mutual shadow is
\begin{equation}
    A_{12}
    =
    \frac{h^2}{2}
    \sum_{i,j,k,l=1}^{2}
    s_i s_j s_k s_l\,
    \Phi(Y_{ik},Z_{jl}),
    \label{eq:parallel_rectangle_shadow_closed}
\end{equation}
where
\begin{multline}
    \Phi(Y,Z)
    =
    Z\sqrt{1+Y^2}
    \arctan\!\left(
        \frac{Z}{\sqrt{1+Y^2}}
    \right)
    + \\
    Y\sqrt{1+Z^2}
    \arctan\!\left(
        \frac{Y}{\sqrt{1+Z^2}}
    \right)
    -
    \frac{1}{2}
    \ln\!\left(1+Y^2+Z^2\right).
    \label{eq:parallel_rectangle_shadow_phi}
\end{multline}
Equation~\eqref{eq:parallel_rectangle_shadow_closed} is used for the
complete mutual shadow of the parallel aperture pairs considered in
this work.

\bibliographystyle{IEEEtran}
\bibliography{reference.bib}